\documentclass{emulateapj}
\usepackage{amsmath}
\usepackage{amssymb}
\usepackage{latexsym}
\usepackage{textcomp}
\usepackage{graphicx}
\usepackage{subfigure}
\usepackage{lscape}
\usepackage{longtable}
\usepackage{apjfonts}

\slugcomment{Received 2007 October 8; accepted 2007 December 20}

\begin{document}

\title{Metallicities and Physical Conditions in Star-forming Galaxies at $z$ $\sim$ 1.0-1.5\altaffilmark{1}}

\shorttitle{STAR-FORMING GALAXIES AT Z$\sim$1.0-1.5}
\shortauthors{LIU ET AL.}

\author{\sc Xin Liu, Alice E. Shapley\altaffilmark{2,3}}
\affil{Department of Astrophysical Sciences, Princeton University,
Peyton Hall -- Ivy Lane, Princeton, NJ 08544}

\author{\sc Alison L. Coil\altaffilmark{4}}
\affil{Steward Observatory, University of Arizona, Tucson, AZ 85721}

\author{\sc Jarle Brinchmann}
\affil{Astrofisica da Universidade do Porto, Rua das Estrelas, 4150-762 Porto, Portugal}

\author{\sc Chung-Pei Ma}
\affil{Department of Astronomy, University of California,
Berkeley, 601 Campbell Hall, Berkeley, CA 94720}

\altaffiltext{1}{Based, in part, on data obtained at the W.M. Keck
Observatory, which is operated as a scientific partnership among the
California Institute of Technology, the University of California, and
NASA, and was made possible by the generous financial support of the W.M.
Keck Foundation.}
\altaffiltext{2}{Alfred P. Sloan Fellow}
\altaffiltext{3}{David and Lucile Packard Fellow}
\altaffiltext{4}{Hubble Fellow}

\begin{abstract}
We present a study of the mass-metallicity ($M-Z$) relation and
H~II region physical conditions in a sample of 20 star-forming
galaxies at $1.0 < z < 1.5$ drawn from the DEEP2 Galaxy Redshift
Survey. We find a correlation between stellar mass and gas-phase
oxygen abundance in the sample and compare it with the one
observed among UV-selected $z \sim 2$ star-forming galaxies and
local objects from the Sloan Digital Sky Survey (SDSS). This
comparison, based on the same empirical abundance indicator,
demonstrates that the zero point of the $M-Z$ relationship evolves
with redshift, in the sense that galaxies at fixed stellar mass
become more metal-rich at lower redshift. Measurements of [O
III]/H$\beta$ and [N II]/H$\alpha$ emission-line ratios show that,
on average, objects in the DEEP2 $1.0 < z < 1.5$ sample are
significantly offset from the excitation sequence observed in
nearby H~II regions and SDSS emission-line galaxies. In order to
fully understand the causes of this offset, which is also observed
in $z\sim2$ star-forming galaxies, we examine in detail the small
fraction of SDSS galaxies that have similar diagnostic ratios to
those of the DEEP2 sample. Some of these galaxies indicate
evidence for AGN and/or shock activity, which may give rise to
their unusual line ratios and contribute to Balmer emission lines
at the level of $\sim 20$\%. Others indicate no evidence for AGN
or shock excitation yet are characterized by higher electron
densities and temperatures, and therefore interstellar gas
pressures, than typical SDSS star-forming galaxies of similar
stellar mass. These anomalous objects also have higher
concentrations and star formation rate surface densities, which
are directly connected to higher interstellar pressure. Higher
star formation rate surface densities, interstellar pressures, and
H~II region ionization parameters may also be common at high
redshift. These effects must be taken into account when using
strong-line indicators to understand the evolution of heavy
elements in galaxies. When such effects are included, the inferred
evolution of the $M-Z$ relation out to $z\sim 2$ is more
significant than previous estimates.
\end{abstract}


\section{Introduction}

The abundance of heavy elements in the H~II regions of galaxies
reflects the past history of star formation and the effects of
inflows and outflows of gas. A characterization of the evolution
of chemical abundances for galaxies of different masses is
therefore essential to a complete model of galaxy formation that
includes the physics of baryons \citep{delucia04,finlator07}.
Important observational constraints for such models come from
determining the scaling relations at different redshifts among
galaxy luminosity, stellar mass, and metallicity, which, for
star-forming galaxies, typically consists of the oxygen abundance.
However, one of the key challenges is to take the observationally
measured quantities, i.e. strong, rest-frame optical emission-line
ratios, and connect them with the physical quantity of interest,
i.e. oxygen abundance.

%-----------------------------------------------------------------------------
\begin{deluxetable*}{lcccccccc}
\tablewidth{0pt} \tabletypesize{\footnotesize}
\tablecaption{Galaxies Observed with Keck~II
NIRSPEC\label{tab:obs}}
%\rotate
\tablehead{
\colhead{~~~~~~~~~~DEEP ID~~~~~~~~~~} &
\colhead{~~~~~R.A. (J2000)~~~~~} &
\colhead{~~~~~Dec. (J2000)~~~~~} &
\colhead{~~~~~$z_{{\rm H}\alpha}$~~~~~} &
\colhead{~~~~~$B$~~~~~} &
\colhead{~~~~~$R$~~~~~} &
\colhead{~~~~~$I$~~~~~} &
\colhead{~~~~~$M_B$~~~~~} &
\colhead{~~~~~$U-B$~~~~~}
}
\startdata
    42044579 \dotfill   &    02 30 43.46   &   00 42 43.60 & 1.0180 &   23.22   &    22.97   &    22.40  &  -21.20 &   0.54 \\
    22046630 \dotfill   &    16 50 13.83   &   35 02 01.78 & 1.0225 &   23.64   &    23.02   &    22.31  &  -21.37 &   0.69 \\
    22046748 \dotfill   &    16 50 14.55   &   35 02 04.31 & 1.0241 &   24.43   &    23.76   &    22.90  &  -20.86 &   0.86 \\
    42044575 \dotfill   &    02 30 44.85   &   00 42 51.33 & 1.0490 &   23.08   &    22.94   &    22.56  &  -21.06 &   0.34 \\
    42010638 \dotfill   &    02 29 08.74   &   00 23 28.40 & 1.3877 &   22.93   &    22.85   &    22.54  &  -22.12 &   0.49 \\
    42010637 \dotfill   &    02 29 08.74   &   00 23 32.87 & 1.3882 &   24.20   &    23.98   &    23.72  &  -20.87 &   0.44 \\
    42021412 \dotfill   &    02 30 44.55   &   00 30 50.73 & 1.3962 &   24.07   &    23.74   &    23.12  &  -21.91 &   0.78 \\
    42021652 \dotfill   &    02 30 44.70   &   00 30 46.19 & 1.3984 &   22.97   &    22.24   &    21.32  &  -24.01 &   1.01 \\
\enddata
\tablecomments{Units of right ascension are hours, minutes, and seconds, and units of declination are degrees, arcminutes and arcseconds.}
\end{deluxetable*}
%-----------------------------------------------------------------------------

In the local universe, \citet{tremonti04} have used a sample of
$\sim$53,000 emission-line galaxies from the Sloan Digital Sky
Survey (SDSS) to investigate the luminosity-metallicity ($L-Z$)
and mass-metallicity ($M-Z$) relationships. For this sample,
metallicities were estimated on the observed spectra of several
strong emission lines, including [O II] $\lambda\lambda$3726,
3729, H$\beta$, [O III]$ \lambda\lambda$5007, 4959, H$\alpha$, [N
II] $\lambda\lambda$6548, 6584, and [S II] $\lambda\lambda$6717,
6731. At increasing redshifts, as the strong rest-frame optical
emission lines shift into the near-IR, metallicities are typically
based on smaller subsets of strong emission lines through the use
of empirically calibrated abundance indicators
\citep[e.g.][]{pp04,pagel79}. Much progress has been made recently
in assembling large samples of star-forming galaxies with
abundance measurements at both intermediate redshift
\citep{kewley02,savaglio05} and at $z>2$ \citep{erb06a}. However,
we have only begun to gather chemical abundance measurements for
galaxies at $z \sim 1-2$ \citep[][hereafter Paper
I,]{maier06,shapley05}. In this work, we continue our efforts to
fill in the gap of chemical abundance measurements during this
important redshift range, which hosts the emergence of the Hubble
sequence of disk and elliptical galaxies \citep{dickinson00}, and
the buildup of a significant fraction of the stellar mass in the
universe \citep{drory05,dickinson03} prior to the decline in
global star formation rate (SFR) density \citep{madau96}.

Chemical abundances for high-redshift galaxies are commonly
estimated using locally calibrated empirical indicators. Yet it is
crucial to recognize the fact that a considerable fraction of the
$z \sim 1$ and $2$ galaxies with measurements of multiple
rest-frame optical emission lines do not follow the local
excitation sequence described by nearby H~II regions and
star-forming galaxies in the diagnostic diagram featuring the [O
III] $\lambda$5007/H$\beta$ and [N II] $\lambda$6584/H$\alpha$
emission-line ratios \citep[Paper I;][]{erb06a}. On average, the
distant galaxies lie offset towards higher [O III]
$\lambda$5007/H$\beta$ and [N II] $\lambda$6584/H$\alpha$,
relative to local galaxies. As discussed in Paper I and
\citet{groves06}, several causes may account for this offset, in
terms of the prevailing physical conditions in the H~II regions of
high-redshift galaxies. The relevant conditions are H~II region
electron density, hardness of the ionizing spectrum, ionization
parameter, the effects of shock excitation, and contributions from
an active galactic nucleus (AGN). It is still unclear which of
these are most important for determining the emergent spectra of
high- redshift galaxies. Understanding this offset in
emission-line ratios is important, not only because it provides
evidence that physical conditions in the high redshift universe
are different from the local ones, but also because the
application of an empirically calibrated abundance indicator to a
set of H~II regions or star-forming galaxies rests on the
assumption that these objects are similar, on average, to those on
which the calibration is based.

In this sense, the current work is also motivated by the
interpretation of the offset in emission-line ratios among distant
galaxies, and an assessment of the reliability of using local
abundance calibrations for high-redshift star-forming galaxies.
Instead of focusing on high-redshift objects, another approach is
to study the properties in a class of nearby objects, which
exhibit similar offset behavior on the emission-line diagnostic
diagram. Unravelling the relations between the physical conditions
and unusual diagnostic line ratios for such objects aids the
understanding of high-redshift galaxies. The SDSS, with its rich
set of photometric and spectroscopic information, provides an
ideal local comparison sample.

In this paper we expand on the analysis presented in Paper I, with
an enlarged sample of DEEP2 star-forming galaxies observed with
NIRSPEC on the Keck II telescope. The larger number of DEEP2
objects with near-IR observations enforces the conclusions drawn
in the previous work. Furthermore, our detailed study of nearby
SDSS objects with similar emission-line diagnostic ratios leads to
a clearer physical explanation of the observed properties of the
DEEP2 galaxies. The DEEP2 sample, near-IR spectroscopic
observations, data reduction, and measurements are described in \S
2. We present the oxygen abundances derived from measurements of
[O III], H$\beta$, H$\alpha$, and [N II] emission lines in both
individual as well as composite spectra in \S 3. The
mass-metallicity relationship and its evolution through cosmic
time are discussed in \S 4. In \S 5 we investigate differences in
$z \sim 1.0-1.5$ H II region physical conditions with respect to
local samples by examining nearby SDSS galaxies with similar
emission-line diagnostic ratios. Finally, we summarize our main
conclusions in \S 6. A cosmology with $\Omega_m = 0.3$,
$\Omega_{\Lambda} = 0.7$, and $h = 0.7$ is assumed throughout.

%---------------------------------------------------------------------------------------------
\section{DEEP2 Sample, Observations, and Data Reduction}

\subsection{DEEP2 Target Sample and Near-IR Spectroscopy}

The high-redshift galaxies presented in this paper are drawn from
the DEEP2 Galaxy Redshift Survey \citep[hereafter
DEEP2;][]{davis03,faber05}, which contains $>30,000$ galaxies with
high-confidence redshifts at $0.7 \leq z \leq 1.5$ down to a
limiting magnitude of $R_{AB}=24.1$. The motivation for our
follow-up near-infrared spectroscopic program, along with detailed
descriptions of the sample selection, optical and near-IR
photometry, and spectroscopy are presented in Paper I. Only a
brief overview is given here.

The new sample contains four galaxies at $z \sim 1.0$ and four
galaxies at $z \sim 1.4$, which, in combination with the pilot
program presented in Paper I, leads to a sample of 20 galaxies in
total. These galaxies are located in fields 2, 3, and 4 of the
DEEP2 survey, at 16, 23, and 2 hr right ascension, respectively.
To probe chemical abundances and H~II region physical conditions,
observations of several strong H~II region emission lines are
required, ideally at least [O II], H$\beta$, [O III], H$\alpha$,
and [N II]. At $z \geq 0.85$, however, the only strong H~II region
emission feature contained in the DEEP2 DEIMOS spectroscopic data
is the [O II] doublet.  Therefore, near-IR spectroscopy is needed
to measure longer wavelength H~II region emission lines at $z \geq
1$. We target two narrow redshift windows within the larger DEEP2
redshift distribution: $0.96 \leq z \leq 1.05$ and $1.36 \leq z
\leq 1.50$, within which it is possible to measure the full set of
H$\beta$, [O III], H$\alpha$, and [N II], in spite of the bright
sky lines and strong atmospheric absorption in the near-IR.

%---------------------------------------------
\begin{figure*}
\epsscale{.8} \plotone{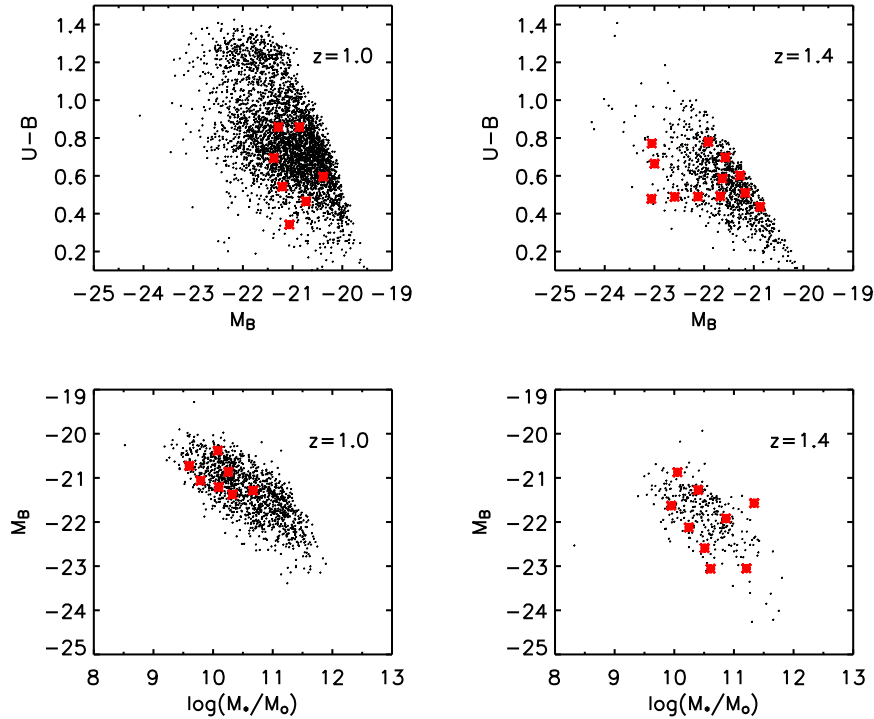} \caption{DEEP2 color-magnitude and
magnitude-stellar mass diagrams. The rest-frame $(U - B)$ vs.
$M_B$ color-magnitude diagrams ({\it top}) and the $M_B$ vs.
stellar mass diagrams ({\it bottom}) are for DEEP2 galaxies at
$0.96 \le z \le 1.05$ ({\it left}) and $1.36 \le z \le 1.50$ ({\it
right}). In each plot, DEEP2 galaxies from both the pilot sample
presented in Paper I as well as objects with new NIRSPEC
observations are indicated with red squares. As shown here, all
NIRSPEC targets were drawn from the ``blue cloud'' of the color
bimodality. \label{fig:colormag}.}
\end{figure*}
%---------------------------------------------

The absolute $B$ magnitude, $M_B$, and stellar mass estimates are
given in Tables \ref{tab:obs} and \ref{tab:emi} for the new
objects and plotted as red squares in the lower panels of Figure
\ref{fig:colormag}, together with the data for the pilot sample.
For all the objects, we use the $M_B$ values from
\citet{willmer06} based on optical data alone and confirm their
good agreement with estimates based on fits to the $BRIK_S$ SEDs
that span through rest-frame $J$ ($I$) band for the objects at $z
\sim 1.0$ ($z \sim 1.4$). Stellar masses for the objects in our
sample are derived with $K_S$-band photometry, following the
procedure outlined in \citet{bundy05}, which assumes a
\citet{chabrier03} stellar initial mass function (IMF). As
discussed in detail in Paper I, the \citet{bundy05} stellar mass
modelling technique agrees well with that used by
\citet{kauffmann03a} for SDSS galaxies, based on both spectral
features and broadband photometry. Stellar masses for galaxies in
the pilot sample have been updated to reflect both the most
current DEEP2 near-IR photometric catalog, and population
synthesis models limiting the stellar population age to be younger
than the age of the Universe at $z\sim 1$. Thus, in a few cases,
the stellar masses differ slightly from those presented in Table
(2) of Paper I. As shown in the lower panels of Figure
(\ref{fig:colormag}), the $z \sim 1.4$ galaxies in our sample span
the full range of absolute $B$ luminosities in the DEEP2 survey,
from $M_{B} \sim$ -20 to -23, while the smaller set of $z \sim
1.0$ galaxies happens to cover the faint end of the luminosity
function. Galaxies in both redshift intervals cover more than an
order of magnitude in stellar mass and therefore should be able to
probe an interesting dynamic range. Since our goal was to study
the emission-line properties of galaxies, all objects in our
sample lie in the blue component of the observed $(U-B)$ color
bimodality in the DEEP2 survey, as shown in the upper panels of
Figure \ref{fig:colormag}.

% ---------------------------------------------
\begin{deluxetable*}{lcccccccccr}
\tabletypesize{\scriptsize}
% \rotate
\tablecaption{Emission Lines and Physical Quantities.\label{tab:emi}}
\tablewidth{0pt}
\tablehead{
     &  &  &  &  &  &
    \multicolumn{2}{c}{12 + log(O/H)} &  &  &  \\
    \cline{7-8} \\
    \colhead{DEEP ID} & \colhead{z$_{{\rm H}\alpha}$} & \colhead{F$_{{\rm H}\beta}$\tablenotemark{a}} &
    \colhead{F$_{{\rm [O III]}\lambda5007}$\tablenotemark{a}} & \colhead{F$_{{\rm H}\alpha}$\tablenotemark{a}} &
    \colhead{F$_{{\rm [N II]}\lambda6584}$\tablenotemark{a}} & \colhead{N2\tablenotemark{b}} &
    \colhead{O3N2\tablenotemark{c}} & \colhead{$L_{H\alpha}$\tablenotemark{d}} &
    \colhead{SFR$_{{\rm H}\alpha}$\tablenotemark{e}}  &
    \colhead{log$(M_{\ast}/M_{\odot})$\tablenotemark{f}}
    }
\startdata
42044579 & 1.0180 & 2.4$\pm$0.4 &  5.4$\pm$0.3 & 12.1$\pm$0.3 & 2.8$\pm$0.3 & 8.54$\pm$0.18 & 8.41$\pm$0.14 & 0.7 &  3  & 10.26$\pm$0.15 \\
22046630 & 1.0225 & 3.7$\pm$0.5 &  8.7$\pm$0.4 & 15.2$\pm$0.3 & 1.8$\pm$0.3 & 8.37$\pm$0.18 & 8.32$\pm$0.14 & 0.8 &  3  & 10.29$\pm$0.16 \\
22046748 & 1.0241 & 2.2$\pm$0.4 &  6.4$\pm$0.3 &  9.6$\pm$0.2 & 2.3$\pm$0.2 & 8.55$\pm$0.18 & 8.39$\pm$0.14 & 0.5 &  2  & 10.28$\pm$0.10 \\
42044575 & 1.0490 & 7.7$\pm$0.7 & 22.9$\pm$0.3 & 23.6$\pm$0.3 & 3.6$\pm$0.3 & 8.43$\pm$0.18 & 8.32$\pm$0.14 & 1.4 &  6  &  9.74$\pm$0.06 \\
42010638 & 1.3877 & 3.8$\pm$0.9 & 17.3$\pm$0.4 & 16.6$\pm$0.5 & 2.8$\pm$0.5 & 8.46$\pm$0.19 & 8.27$\pm$0.15 & 2.0 &  9  & 10.21$\pm$0.06 \\
42010637 & 1.3882 & 3.3$\pm$0.9 &  3.7$\pm$0.3 &  6.4$\pm$0.4 & 2.9$\pm$0.4 & 8.70$\pm$0.18 & 8.60$\pm$0.15 & 0.8 &  4  & 9.96$\pm$0.12 \\
42021412 & 1.3962 & 7.1$\pm$0.8 &  $>$ 2.0     & 12.7$\pm$0.3 & 2.9$\pm$0.3 & 8.54$\pm$0.18 &  $<$ 8.70     & 1.5 &  7  & 10.89$\pm$0.13 \\
42021652\tablenotemark{g} & 1.3984 & 4.6$\pm$0.5 &  9.7$\pm$0.5 &  9.5$\pm$0.3 & 1.2$\pm$0.3 & 8.38$\pm$0.19 & 8.33$\pm$0.15 & 1.1 & 5 & ... \\
\enddata
%%%%%%%
\tablenotetext{a}{Emission-line flux and random error in units of $10^{-17}$ ergs s$^{-1}$ cm$^{-2}$.} 
\tablenotetext{b}{Oxygen abundance deduced from the N2 relationship presented in \citet{pp04}.} 
\tablenotetext{c}{Oxygen abundance deduced from the O3N2 relationship presented in \citet{pp04}.}
\tablenotetext{d}{H$\alpha$ luminosity in units of $10^{42}$ ergs s$^{-1}$.} 
\tablenotetext{e}{Star formation rate in units of $M_{\odot}$ yr$^{-1}$, calculated from $L_{{\rm H}\alpha}$ using the calibration of \citet{kennicutt98}, and divided by a factor of 1.8 to convert to a \citet{chabrier03} IMF from the Salpeter IMF assumed by \citet{kennicutt98}. Note that SFRs have not been corrected for dust extinction or aperture effects, which may amount to a factor of $2$ difference \citep{erb06c}.}
\tablenotetext{f}{Stellar mass and uncertainty estimated using the methods described in Bundy et al. (2005), and assuming a \citet{chabrier03} IMF.} \tablenotetext{g}{This object has a double morphology. The separation between the two components is about 0.9$^{''}$, which corresponds to $\sim$8 kpc at $z = 1.3984$. We measured line fluxes for the emission-line-dominated component, but do not have a robust estimate of the corresponding stellar mass. In the DEEP2 photometry the two components were counted as one source, and the resulting stellar-mass estimate has contribution from both components. We therefore do not include this stellar mass estimate in our sample.}
\end{deluxetable*}

The near-IR spectra were obtained on 2005 September 17 and 18 with
the NIRSPEC spectrograph \citep{mclean98} on the Keck~II
telescope. Over the range of redshifts of the galaxies presented
here, two filter setups are required to measure the full set of
H$\beta$, [O III], H$\alpha$, and [N II]. For objects at $z \sim
1.4$, the NIRSPEC 5 filter (similar to $H$ band) is used to
observe H$\alpha$ and [N II], whereas the NIRSPEC 3 filter
(similar to $J$ band) is used for H$\beta$ and [O III]. For
objects at $z \sim 1.0$, the NIRSPEC 3 filter is used to observe
H$\alpha$ and [N II], whereas the NIRSPEC 1 filter ($\Delta\lambda
=$ 0.95-1.10 $\mu$m) is used for H$\beta$ and [O III]. All targets
were observed for 3$\times$900 s in each filter with a
0.76$^{''}$$\times$42$^{''}$ long slit. The spectral resolution
determined from sky lines is $\sim10 {\rm \AA}$ for all four
NIRSPEC filters used here. Photometric conditions and seeing were
variable throughout both nights, with seeing ranging from
0.5$^{''}$ to 0.7$^{''}$ in the near-IR. In order to enhance the
long-slit observing efficiency, we targeted two galaxies
simultaneously by placing them both on the slit.

We observed a total of 10 DEEP2 galaxies, successfully measuring
the full set of H$\beta$, [O III], H$\alpha$, and [N II] for eight
out of 10. For the remaining pair, we only detected H$\alpha$ in
the $H$ band, but no H$\beta$ nor [O III] in the $J$-band
exposures, in which the background in between sky lines was
characterized by a significantly higher level of continuum than
usual. This anomalous background is likely due to an increased
contribution from clouds, which may have affected both $H$- and
$J$-band observations of the pair. Since a clear measurement was
not obtained for these two objects, due to variable weather
conditions, we exclude them from our study. The object, 42021652,
has a double morphology, with one component dominated by emission
lines with a weak continuum, and another component dominated by
strong continuum, with only weak emission lines at roughly the
same redshift. The separation between the two components on the
sky is $\sim$0.9$^{''}$, which corresponds to $\sim$8 kpc at $z =
1.3984$, perhaps indicative of a merger event. This interpretation
is supported by the small velocity difference of $\Delta v\sim
125$~km s$^{-1}$ between the two components. We measure line
fluxes for the component dominated by emission lines, since it
provides a more robust estimate of line ratios. Deblended optical
and near-infrared magnitudes would be required to obtain robust
stellar masses for the individual components. However, in the
DEEP2 photometry the two components were counted as one source
since they are too close to be deblended and the stellar mass has
contribution from both of them. For now, we only include flux
measurements of the emission-line component for the
diagnostic-line-ratio analysis but do not include this object in
the mass-metallicity studies. A summary of the observations
including target coordinates, redshifts, and optical and near-IR
photometry is given in Table \ref{tab:obs}.

\subsection{Data Reduction and Optimal Background Subtraction}

Data reduction was performed with a similar procedure to the one
described in Paper I and \citet{erb03}, with the exception of an
improved background subtraction method applied to the
two-dimensional galaxy spectral images \citep[][private
communication]{kelson03,becker06}.
In the custom NIRSPEC long-slit reduction package written by D. G.
Becker (2006, private communication), optimal background
subtraction is performed on the unrectified science frames. First,
a transformation is calculated between CCD ($x$, $y$) coordinates
and those of slit position and wavelength, using the
wavelength-dependent traces of bright standard stars and the
spatially dependent curves of bright sky lines. Then a
two-dimensional model of the sky background is constructed as a
function of slit position and wavelength, using a low-order
polynomial in the slit-position dimension, and a $b$-spline
function in the wavelength dimension. This two-dimensional model
is iteratively fit in the differenced frame of adjacent science
exposures and subtracted from the unrectified data. After
background subtraction, cosmic rays were removed from each
exposure, which was then rotated, cut out along the slit, and
rectified. Finally, all background-subtracted, rectified exposures
of a given science target were combined in two dimensions. This
new approach to reducing NIRSPEC spectra results in fewer
artifacts around bright sky lines and cosmic rays, which are
commonly introduced when rectification is performed before sky
subtraction and cosmic-ray zapping.
One-dimensional spectra, along with error spectra, were then
extracted and flux- calibrated using A-star observations,
according to the procedure described in Paper I and \citet{erb03}.

%%%%%%%%---------------------------------------------------------------------------------------------

\subsection{Measurements and Physical Quantities}

% ---------------------------------------------
\begin{figure*}
\epsscale{1} \plotone{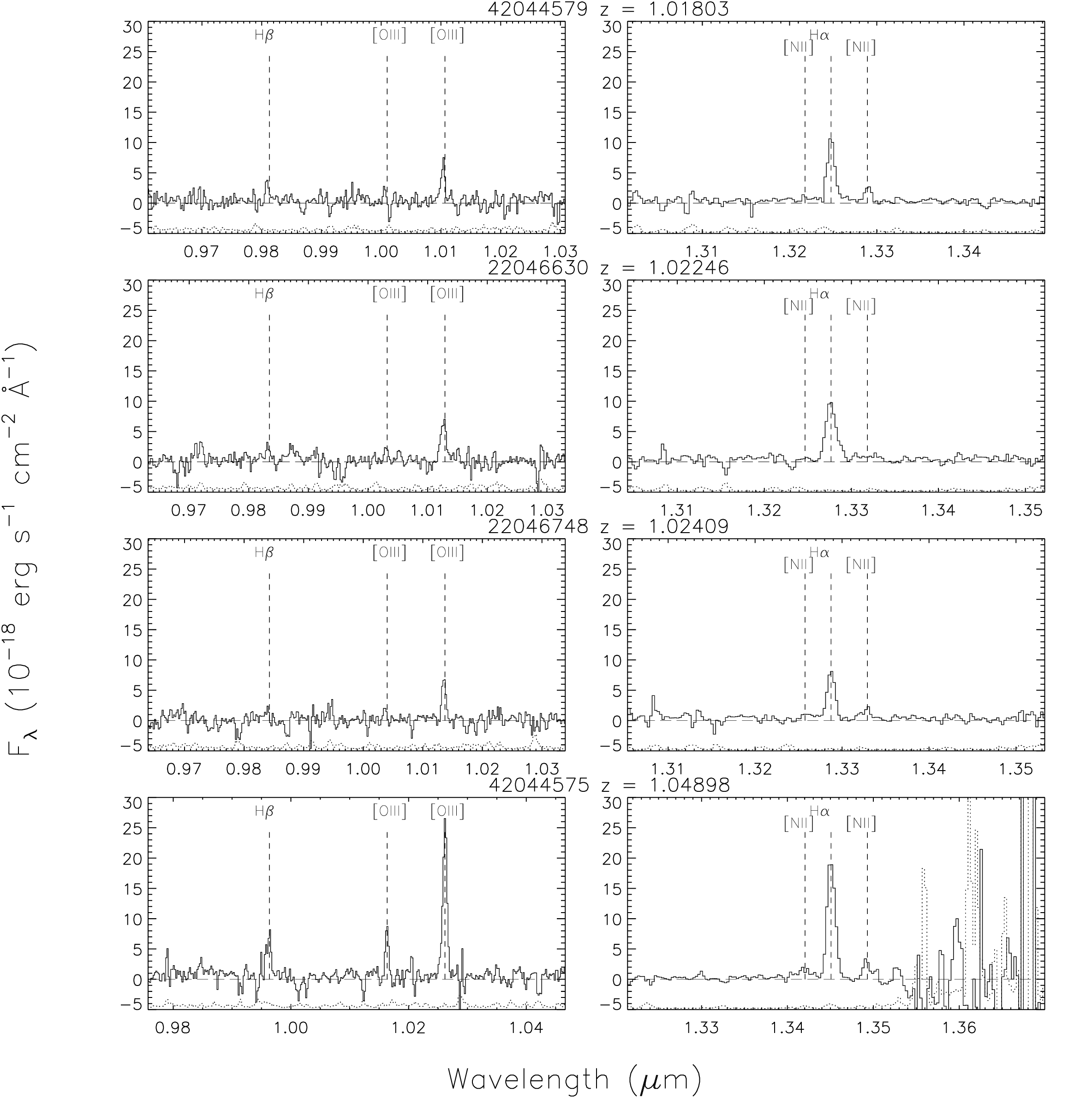} \caption{NIRSPEC spectra of DEEP2
galaxies in our new sample at $z\sim 1.0$. H$\beta$ and [O III]
are observed in the NIRSPEC 1 filter, with H$\alpha$ and [N II] in
the NIRSPEC 3 filter (similar to the $J$ band). The 1 $\sigma$
error spectra are shown as dotted lines, offset vertically by $-5
\times 10^{-18}$ ergs s$^{-1}$ cm$^{-2}$ \AA$^{-1}$ for
clarity.\label{fig:spec1}} \epsscale{1.}
\end{figure*}

% ---------------------------------------------

\begin{figure*}
\epsscale{1.} \plotone{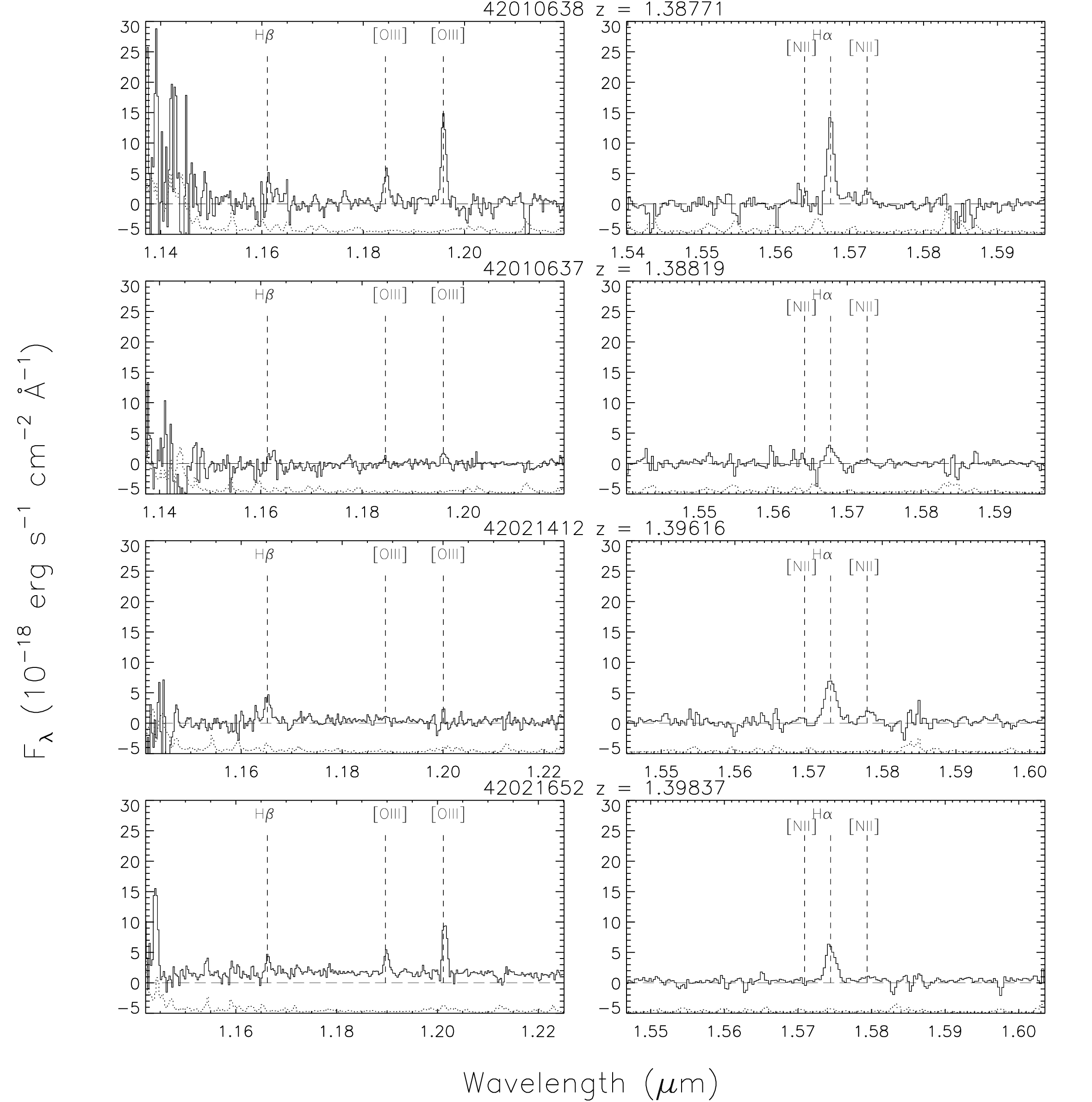} \caption{NIRSPEC spectra of DEEP2
galaxies in our new sample at $z\sim 1.4$. H$\beta$ and [O III]
are observed in the NIRSPEC 3 filter, with H$\alpha$ and [N II] in
the NIRSPEC 5 filter (similar to the $H$ band). The 1 $\sigma$
error spectra are shown as dotted lines, offset vertically by $-5
\times 10^{-18}$ ergs s$^{-1}$ cm$^{-2}$ \AA$^{-1}$ for
clarity.\label{fig:spec2}} \epsscale{1.}
\end{figure*}

One-dimensional, flux-calibrated NIRSPEC spectra along with the 1
$\sigma$ error spectra of galaxies in our new sample are shown in
Figures \ref{fig:spec1} and \ref{fig:spec2}. Emission-line fluxes
and uncertainties measured from the one-dimensional spectra are
given in Table \ref{tab:emi}. H$\alpha$ and [N II] $\lambda$6584
emission-line fluxes were determined by first fitting a Gaussian
profile to the H$\alpha$ feature to obtain the redshift and FWHM,
and using these values to constrain the fit to the [N II] emission
line. This method is based on the assumption that the H$\alpha$
and [N II] lines have exactly the same redshift and FWHM, with the
H$\alpha$ line having a higher signal-to-noise ratio (S/N). For
most of the objects in our sample, the [N II] $\lambda$6548 line
was too faint to measure. [O III] $\lambda$5007 and H$\beta$
fluxes were determined with independent fits. In most cases,
redshifts from H$\alpha$, [O III] $\lambda$5007, and H$\beta$
agree to within $\Delta z = 0.0004$ ($\Delta v = 50-60$ km
s$^{-1}$ at $z=1.0-1.4$). For the object 42021412, [O III]
$\lambda$5007 lies on top of a bright sky line, and only a lower
limit is given.

SFRs inferred from H$\alpha$ luminosities using the calibration of
\citet{kennicutt98} are shown in Table (\ref{tab:emi}). The
results have been converted from the Salpeter IMF used by
\citet{kennicutt98} to a \citet{chabrier03} IMF by dividing the
results by a factor of $1.8$. In our whole sample of 20 galaxies,
the H$\alpha$ fluxes range from $5.6 \times 10^{-17}$ to $2.4
\times 10^{-16}$ erg s$^{-1}$ cm$^{-2}$. The mean H$\alpha$ flux
for the sample at $z \sim 1.0$ ($z \sim 1.4$) is $1.3 \times
10^{-16}$ ($1.2 \times 10^{-16}$), corresponding to a star
formation rate of 3 (6) $M_{\odot}$ yr$^{-1}$, uncorrected for
dust extinction or aperture effects, which may amount to a factor
of $2$ difference \citep{erb06c}. Note that these characteristic
H$\alpha$ star formation rates, after being corrected for aperture
effects, would be significantly higher than those of local
galaxies in the SDSS sample of \citet{tremonti04} and the $\langle
z \rangle = 0.4$ TKRS subsample of \citet{kobulnicky04}, even
before correction for dust extinction. We also note that the mean
specific SFR for both $z\sim 1.0$ and $z \sim 1.4$ samples is
$\log((SFR/M_{\star})\mbox{ yr}^{-1})=-9.7$.

As discussed in Paper I, absolute line flux measurements suffer
from several sources of systematic error, which can amount to at
least a $\sim25\%$ uncertainty \citep{erb03}. This level of
uncertainty is present even under photometric conditions, which
may not have applied through the full extent of our observations.
For the remainder of the discussion we therefore focus on the
measured line {\it ratios}, [N II] $\lambda$6584/H$\alpha$ and [O
III] $\lambda$5007/H$\beta$, which are not only unaffected by
uncertainties in flux calibration and other systematics but also
relatively free from the effects of dust extinction, due to the
close wavelength spacing of the lines in each ratio. Hereafter, we
use ``[N II]/H$\alpha$" to refer to the measured emission-line
flux ratio between [N II] $\lambda$6584 and H$\alpha$, and ``[O
III]/H$\beta$" for that between [O III] $\lambda$5007 and
H$\beta$.

\section{The Oxygen Abundance}

H II region metallicity is an important probe of galaxy formation
and evolution, as it represents the integrated products of past
star formation, modulated by the inflow and outflow of gas. Oxygen
abundance is often used as a proxy for metallicity since oxygen
makes up about half of the metal content of the interstellar
medium and exhibits strong emission lines from multiple ionization
states in the rest-frame optical that are easy to measure. For
comparison, we use the solar oxygen abundance expressed as 12 +
log(O/H) = 8.66 \citep{allende02,asplund04}.

The most robust way to estimate the oxygen abundance is the
so-called direct $T_e$ method, based on the measurement of the
temperature-sensitive ratio of auroral and nebular emission lines.
However, in distant galaxies the auroral lines are almost always
undetectable \citep[but see][]{hoyos05,kakazu07} since they become
extremely weak at metallicities above $\sim$0.5 solar. Even at
lower metallicities, the auroral lines are typically beyond the
reach of the low S/N typical of the spectra of distant galaxies.
For distant star-forming galaxies, therefore, measuring strong
emission-line ratios is the only viable way of obtaining the H II
region gas-phase oxygen abundance \citep{kobulnicky99,pettini01}.

Given our NIRSPEC data set, and our desire to avoid the systematic
uncertainties entailed in adding [O II] line fluxes obtained with
the DEIMOS spectrograph (without real-time flux-calibration) to [O
III] fluxes obtained with NIRSPEC, we focus on two strong-line
ratios as indicators for the oxygen abundance: N2$\equiv$ log([N
II]/H$\alpha$) and O3N2$\equiv$ log\{([O III]/H$\beta$)/([N
II]/H$\alpha$)\}. These indicators have been calibrated by
\citet{pp04} using local H II regions, most of which have direct
$T_e$ abundance determinations. The sensitivity of the indicators
to oxygen abundance, as well as their limitations, have been
discussed in \citet{pp04} and Paper~I.  Absolute estimates of
metal abundances are quite uncertain, as abundances determined
with different indicators or with different calibrations of the
same indicator may have substantial biases or discrepancies
\citep[e.g.][]{kennicutt03,kobulnicky04}. We therefore emphasize
relative abundances determined with the same method, using the
same calibration.

The N2 indicator, pointed out by several works
\citep{storchi94,raimann00,denicolo02}, is related to the oxygen
abundance via
\begin{equation}
{\rm 12 + log(O/H) = 8.90 + 0.57 \times} {\rm N2},
\end{equation}
which is valid for ${\rm7.50 < 12 + log(O/H) < 8.75}$, with a 1
$\sigma$ scatter of $\pm0.18$ dex \citep{pp04}. It has been used
by \citet{erb06a} to estimate oxygen abundances for UV-selected $z
\sim 2$ galaxies. Note that the N2 indicator is not sensitive to
increasing oxygen abundance above roughly solar metallicity, as
shown with photoionization models \citep{kewley02}. Thus, for a
subset of 12 galaxies in our sample with measurements of the full
set of H$\beta$, [O III], H$\alpha$, and [N II], we also use the
O3N2 indicator introduced by \citet{alloin79}, which is expected
to be particularly useful at solar and super-solar metallicities
where [N II] saturates but the strength of [O III] continues to
decrease with increasing metallicity. Using the same calibration
sample, \citet{pp04} show that O3N2 is related to the oxygen
abundance via
\begin{equation}
{\rm 12 + log(O/H) = 8.73 - 0.32 \times} {\rm O3N2},
\end{equation}
which is valid for ${\rm8.12 < 12 + log(O/H) < 9.05}$, with a 1
$\sigma$ scatter of $\pm0.14$ dex. Table \ref{tab:emi} lists
oxygen abundances derived using these two indicators. The errors
on the oxygen abundances are dominated by the systematic
uncertainties in the calibrations of the indicators.

\subsection{Composite Spectra}

% ---------------------------------------------
\begin{figure*}
\epsscale{1.} \plotone{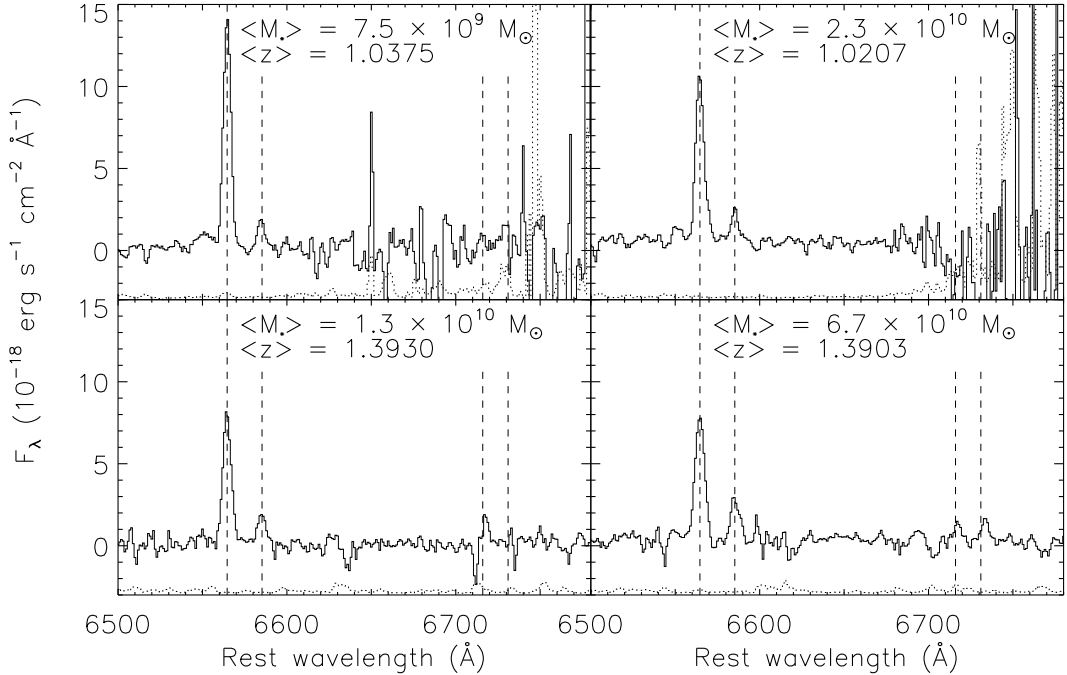} \caption{Composite NIRSPEC spectra
of the 18 $z\sim 1.0-1.5$ galaxies with both stellar mass
estimation as well as [N II] and H$\alpha$ measurement in our
sample, divided separately at $z \sim 1.0$ and $z \sim 1.4$. The 1
$\sigma$ error spectra are shown as dotted lines, offset
vertically by $-3 \times 10^{-18}$ ergs s$^{-1}$ cm$^{-2}$
\AA$^{-1}$ for clarity. The spectra are labeled with the mean
stellar mass from each bin, and the H$\alpha$, [N II] and [S II]
lines are marked by dashed lines. Note that the spectra near the
density-sensitive [S II] lines are very noisy, due to the large
dispersion of the flux per count from near-IR standard star
calibration, caused by low efficiency near the filter
edge.\label{fig:compspec:n2}}
\end{figure*}

\begin{figure*}
\epsscale{.95} \plotone{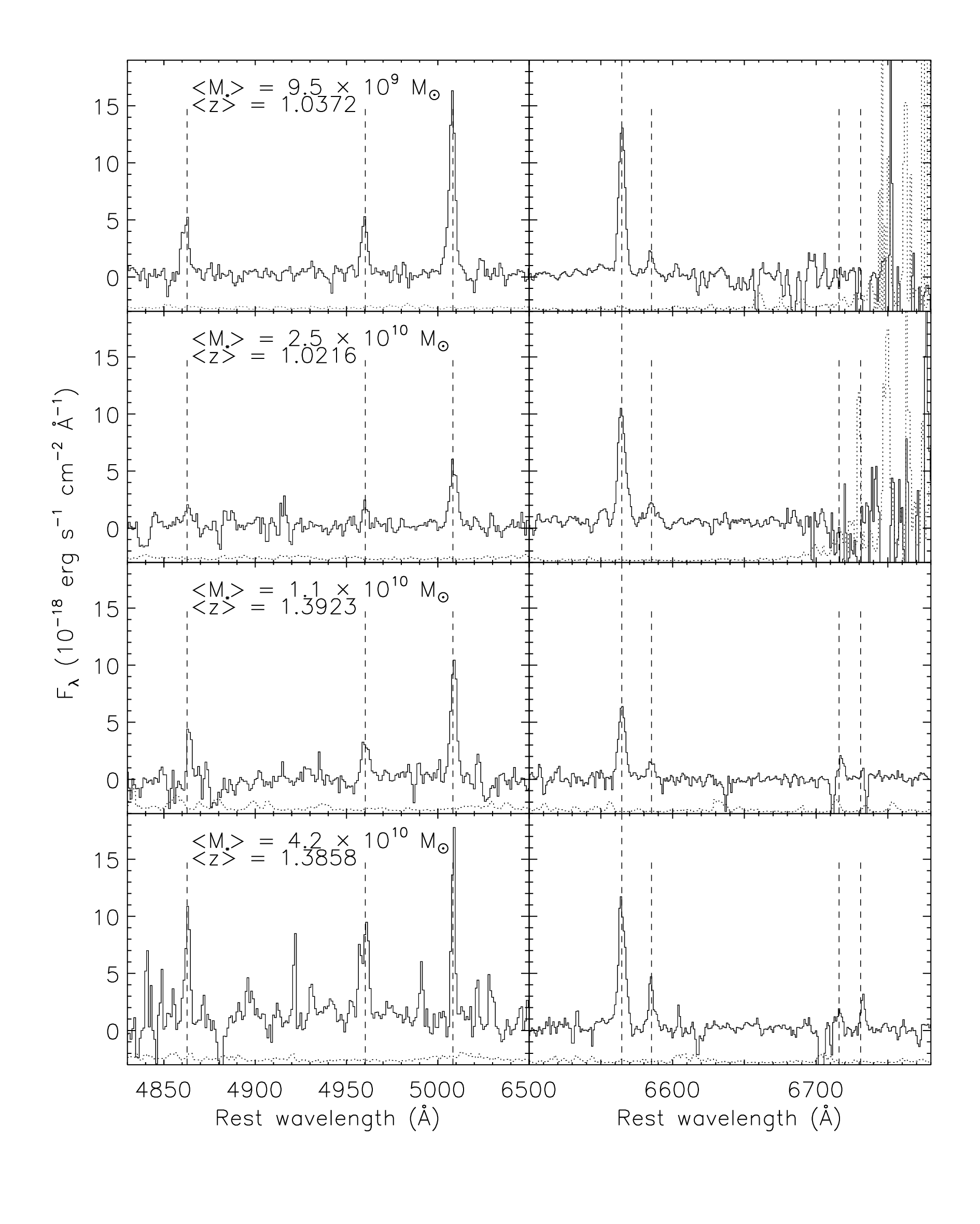} \caption{Composite NIRSPEC spectra
of the 12 $z\sim 1-1.5$ galaxies with both stellar mass estimation
as well as all-four-line measurement in our sample, divided
separately at $z \sim 1.0$ and $z \sim 1.4$. The 1 $\sigma$ error
spectra are shown as dotted lines, offset vertically by $-3 \times
10^{-18}$ ergs s$^{-1}$ cm$^{-2}$ \AA$^{-1}$ for clarity. The
spectra are labeled with the mean stellar mass from each bin, and
the H$\beta$, [O III], H$\alpha$, [N II] and [S II] lines are
marked by dashed lines.\label{fig:compspec:o3n2}}
\end{figure*}

Relative, average abundances determined from composite spectra can
be more accurately determined than those from individual spectra.
As discussed by \citet{erb06a}, making a composite spectrum not
only reduces the uncertainties associated with the strong-line
calibration by a factor $N^{1/2}$, where $N$ is the number of
objects included in the composite spectrum, but also enhances the
spectrum S/N since sky lines generally lie at different
wavelengths for spectra at different redshifts. In addition, one
of our goals is to determine the average properties of subgroups
of galaxies in our sample. For the subset of 18 galaxies in our
sample with H$\alpha$ and [N II] measurements, as well as stellar
mass estimates, we divide the sample into four bins by stellar
mass, with two bins at $z \sim 1.0$ and two bins at $z \sim 1.4$.
For the subset of 12 galaxies with measurements of not only
H$\alpha$ and [N II], but also H$\beta$ and [O III], we also
divide the sample into four bins by stellar mass with two bins
each at $z \sim 1.0$ and at $z \sim 1.4$.

To make the composite spectra, we first shift the individual
one-dimensional flux-calibrated spectra into the rest frame and
then combine them by generating the median spectrum, which
preserves the relative fluxes of the emission features
\citep{vanden01}. We use N2 and N2+O3 composite spectra to refer
to the composites with H$\alpha$ and [N II], and those with all
four lines, respectively. The N2 and N2+O3 composite spectra,
labelled with mean stellar mass in each bin, are shown in Figures
\ref{fig:compspec:n2} and \ref{fig:compspec:o3n2}. The
corresponding emission-line flux ratios along with uncertainties
measured from the composite spectra, as well as the inferred
oxygen abundances, are listed in Tables \ref{tab:comp:n2} and
\ref{tab:comp:o3n2}. The listed errors in 12 + log(O/H) include
the uncertainties from the propagation of emission-line flux
measurements, as well as the systematic scatter from the
strong-line calibration.
As shown in paper I, the systematic discrepancies between the N2-
and O3N2-based abundances are mainly due to the fact that, on
average, DEEP2 galaxies are offset from the excitation sequence
formed by local H II regions and star-forming galaxies. We discuss
this issue in detail in \S \ref{sec:offset}.

\section{The Mass-Metallicity Relation}\label{sec:mz_relation}

\begin{figure*}
\epsscale{1.} \plotone{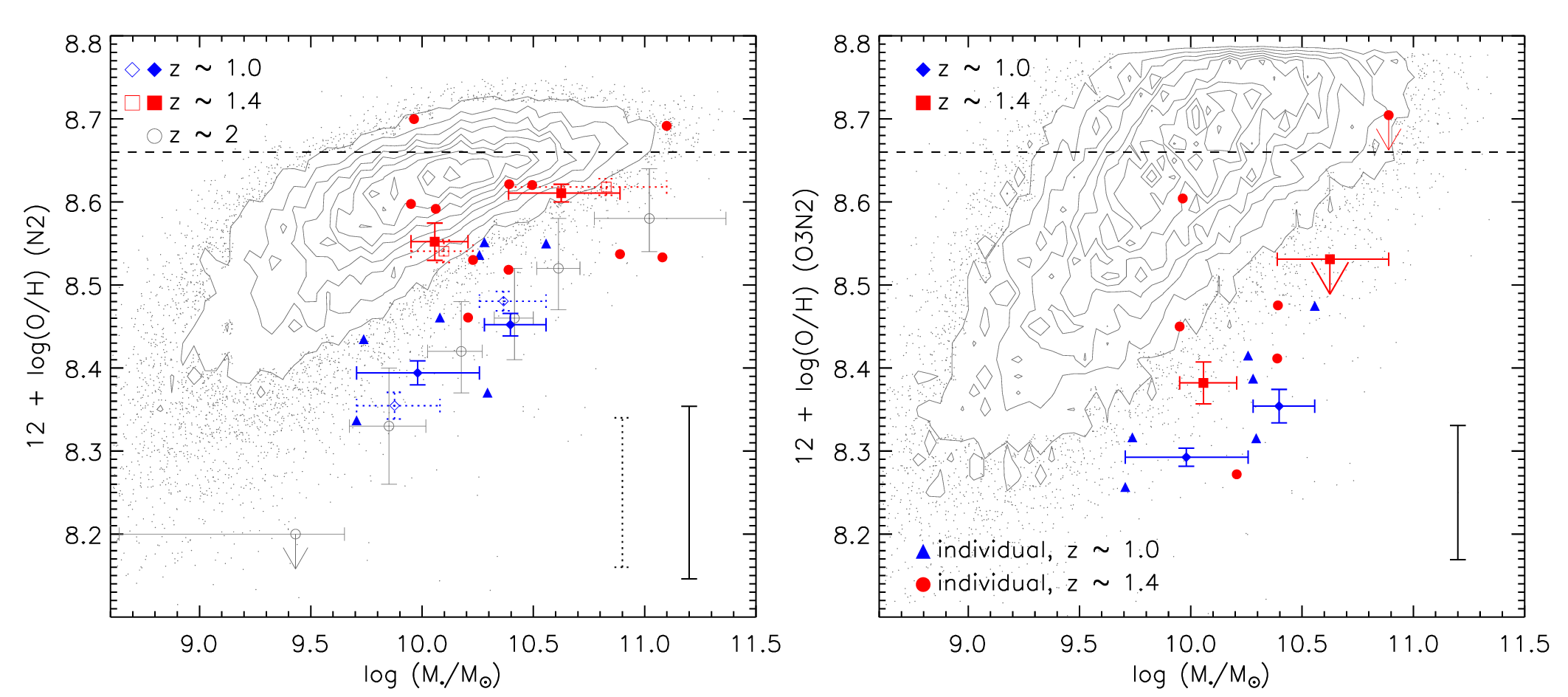} \caption{Mass-metallicity relation
observed at $z \sim 1.0$ and $z \sim 1.4$. In the left panel,
metallicities are inferred from the N2 indicator, while the right
panel shows metallicities estimated from the O3N2 indicator. In
both plots, open (filled) diamonds and squares are used for N2
(N2+O3) composite spectra at $z \sim 1.0$ and $z \sim 1.4$,
respectively (see section 3.1). Errors show the uncertainties
propagated from the strong-line ratio measurement, while the
systematic uncertainties from the strong-line method calibration
\citep{pp04} are shown in the lower right corner of each plot as
dotted (solid) error bars for N2 (N2+O3) composite spectra. The
systematic uncertainties from the calibration are reduced by a
factor $N^{1/2}$, where $N$ is the number of individual spectra
included in each composite. For stellar masses, the horizontal
bars give the mass range in each bin. Data points for individual
objects are shown as filled triangles and circles for $z \sim 1.0$
and $z \sim 1.4$, respectively. Associated error bars are listed
in Table 2 in both this paper and in Paper~I. For comparison,
metallicities as a function of stellar mass are also shown, for
both local SDSS galaxies ({\it grey contours and dots}) and the
\citet{erb06a} $z \sim 2$ sample ({\it grey open circles}, {\it
left panel}). Note that here and throughout, we use contours and
dots to show SDSS objects. On each such applicable plot, SDSS data
points were mapped onto 10 evenly spaced levels according to
number density, where objects on the lowest level are denoted by
dots while other levels are presented by contours. Solar
metallicity is marked with a horizontal dashed line. It can be
seen that the N2 indicator saturates near the solar abundance.
\label{fig:massz}}
\end{figure*}

The redshift evolution of the luminosity-metallicity and
mass-metallicity relations provides important constraints on
models of galaxy evolution. A correlation between gas-phase
metallicity and stellar mass can be explained by either the
tendency of lower mass galaxies to have larger gas fractions and
lower star formation efficiencies
\citep{mcgaugh97,bell00,kobulnicky03}, or the preferential loss of
metals from galaxies with shallow potential wells by
galactic-scale winds \citep{larson74}. In the local universe,
strong correlations between rest-frame optical luminosity and the
degree of chemical enrichment have been observed in both
star-forming and early-type galaxies
\citep{garnett87,brodie91,tremonti04}. The correlation has also
been observed in intermediate- and high- redshift samples
\citep{kobulnicky03,lilly03,kobulnicky04,erb06a}, although caution
must be taken when comparing samples with metallicities determined
from different methods. Physically, the correlation between
stellar mass and metallicity is more fundamental than that between
luminosity and metallicity \citep[Paper I;][]{tremonti04,erb06a}.
We therefore focus on the mass-metallicity relation in the
following discussion.

The left panel of Figure \ref{fig:massz} shows the average
metallicity of the galaxies in each mass bin determined from the
N2 composite spectra plotted against their average stellar mass at
$z \sim 1.0$ ({\it open diamonds}) and at $z \sim 1.4$ ({\it open
squares}). Although our sample is still small, we do see evidence
for mass-metallicity relations at both $z \sim 1.0$ and $z \sim
1.4$. These trends are also present when we examine the
metallicities and stellar masses for individual objects, which are
plotted in the figure as well. For comparison, the local SDSS
galaxies discussed by \citet{tremonti04} are denoted by contours
and dots\footnote{Note that here and throughout, we use a
combination of contours and dots to indicate SDSS objects. On each
such applicable plot, SDSS data points were mapped onto 10 evenly
spaced levels according to surface density, where objects on the
lowest level are denoted by dots while other levels are presented
by contours.} and the $z \sim 2$ \citet{erb06a} sample as open
circles. Metallicities for SDSS galaxies were calculated using the
same strong-line indicator that was applied to the DEEP2 galaxies
and not the Bayesian O/H estimate from \citet{tremonti04}.

At fixed stellar mass, the metallicities of our $z \sim 1.0-1.5$
sample as a whole are lower than those of local galaxies yet
higher than those of the $z \sim 2$ sample. However, there is
evidence for a reverse trend between the subsets of our sample at
$z \sim 1.0$ and at $z \sim 1.4$. In Paper I, this difference was
attributed to the fact that the $z \sim 1.0$ sample was on average
fainter and less massive than the $z\sim 1.4$ sample. With a
larger sample, however, we find that the higher mass $z \sim 1.0$
bin does have lower metallicity than the lower mass $z \sim 1.4$
bin. Differences in outflow or inflow rate of unenriched gas at $z
\sim 1.0$ and at $z \sim 1.4$ could give rise to this trend.
However, the interval in cosmic time between $z\sim 1.4$ and $1.0$
is small enough that typical gas inflow rates at fixed stellar
mass, and the corresponding star formation and outflow rates, will
not significantly evolve. Therefore, this explanation is not a
likely cause of the reverse trend. A different average degree of
dust reddening at $z \sim 1.0$ and at $z \sim 1.4$ is also not a
likely cause, since the N2 indicator is based on emission lines
with very close spacing in wavelength. On the other hand, if the
$z \sim 1.0$ galaxies have systematically different physical
conditions or less significant contributions from AGN activity
relative to the $z \sim 1.4$ objects, metallicities estimated with
the same calibration would be systematically biased between the
two samples in such a way to produce the observed trend. As
discussed in \S \ref{sec:offset}, we propose that the most likely
cause for the reverse trend is this difference in H II region
physical conditions. Since the systematic uncertainty from
strong-line calibration is large, and our sample is still too
small to draw any solid conclusion, it will become feasible to
clarify this issue only when a statistically large enough sample
is assembled, and both the high- and low- mass ends are spanned at
$z \sim 1.0$ as well as at $z \sim 1.4$.

We also plot the mass-metallicity relation from the N2+O3
composite spectra in Figure \ref{fig:massz}, where the left panel
shows metallicities determined from the N2 indicator, while the
right panel shows those determined from the O3N2 indicator, at
both $z \sim 1.0$ ({\it filled diamonds}) and $z \sim 1.4$ ({\it
filled squares}). The O3N2-based abundances are systematically
lower than those based on N2. As suggested in Paper I and
discussed in detail in \S \ref{sec:offset}, these systematic
discrepancies between N2- and O3N2-based abundances are due to the
fact that DEEP2 galaxies depart from the local H~II region
excitation sequence. In addition, the reverse trend between $z
\sim 1.0$ and $1.4$ in metallicity estimated from O3N2 is much
less significant than the one in metallicity estimated from N2. We
return to this issue as well in \S \ref{sec:offset}. Despite these
discrepancies, the overall correlation between average stellar
mass and metallicity observed among the N2 composite spectra is
still present for the N2+O3 spectra. This is evidence that the
correlation is insensitive to the spectrum of any particular
object, as it is robust to analyses using different binning
schemes. At the lower mass end ($M_{\star} \sim 8\times10^{9}$
$M_{\sun}$), the average metallicity of $z \sim 1.0-1.5$ galaxies
based on N2 is at least 0.22 dex lower than the local typical
value. Since the N2 indicator saturates near solar abundance, as
discussed in \citet{erb06a}, it is difficult to determine the true
metallicity offset between two samples at different redshifts
using this indicator. We can also determine the offset from the
O3N2-based abundances, particularly near the solar abundance.
Based on O3N2 abundances, the metallicity offset between our DEEP2
objects and the local SDSS sample is at least 0.21 dex at the
high-mass end ($M_{\star} \sim 5\times10^{10}$ $M_{\sun}$).
However, as we discuss in \S \ref{sec:offset}, these estimates
based on strong-line indicators may still be subject to systematic
uncertainties from using the calibration of H II regions with
significantly different physical properties.

% ---------------------------------------------
\begin{deluxetable}{lccccc}
\tabletypesize{\scriptsize}
\tablecaption{Oxygen Abundances from N2 composite spectra.\label{tab:comp:n2}}
\tablewidth{0pt}
\tablehead
{
\colhead{~~~~~Bin~~~~~} & 
\colhead{N\tablenotemark{a}} & 
\colhead{$\langle {\rm z}_{{\rm H}\alpha} \rangle$\tablenotemark{b}} & 
\colhead{log$(M_{\ast}/M_{\odot})$\tablenotemark{c}} & 
\colhead{N2\tablenotemark{d}} &
\colhead{12 + log(O/H)\tablenotemark{e}}
}
\startdata
1\dotfill & 3 & 1.0375 & $9.88^{+0.10}_{-0.13}$   & -0.96$\pm$0.03 & 8.35$\pm$0.11 \\
2\dotfill & 4 & 1.0207 & $10.37^{+0.05}_{-0.06}$  & -0.74$\pm$0.02 & 8.48$\pm$0.09 \\
3\dotfill & 5 & 1.3930 & $10.10^{+0.06}_{-0.06}$  & -0.63$\pm$0.02 & 8.54$\pm$0.09 \\
4\dotfill & 6 & 1.3903 & $10.83^{+0.05}_{-0.05}$  & -0.49$\pm$0.02 & 8.62$\pm$0.08 \\
\enddata
\tablenotetext{a}{Number of objects contained in each bin.}
\tablenotetext{b}{Mean redshift for each bin.}
\tablenotetext{c}{Mean stellar mass and uncertainty from error propagation.}
\tablenotetext{d}{N2 $\equiv$ log([N II]$\lambda6584$/H$\alpha$).}
\tablenotetext{e}{Oxygen abundance deduced from the N2 relationship presented in \citet{pp04}.}
\end{deluxetable}
% ---------------------------------------------

%---------------------------------------
\begin{deluxetable*}{lccccccc}
\centering
\tabletypesize{\scriptsize}
\tablecaption{Oxygen Abundances from N2+O3 composite spectra.\label{tab:comp:o3n2}}
\tablewidth{0pt}
\tablehead
{
\colhead{~~~~~~~~~~Bin~~~~~~~~~~} & 
\colhead{~~~N~~~} & 
\colhead{~~~$\langle {\rm z}_{{\rm H}\alpha} \rangle$~~~} &
\colhead{~~~log$(M_{\ast}/M_{\odot})$~~~}  & 
\colhead{~~~N2~~~} & 
\colhead{~~~O3N2\tablenotemark{a}~~~}  &
\colhead{~~~[12 + log(O/H)]$_{N2}$\tablenotemark{b}~~~} & 
\colhead{~~~[12 + log(O/H)]$_{O3N2}$\tablenotemark{c}~~~}
}
\startdata
1\dotfill & 3 & 1.0372 &  $9.98^{+0.09}_{-0.11}$  & -0.89$\pm$0.03 &  1.37$\pm$0.03 & 8.39$\pm$0.10 & 8.29$\pm$0.08 \\
2\dotfill & 3 & 1.0216 &  $10.40^{+0.06}_{-0.06}$ & -0.79$\pm$0.02 &  1.17$\pm$0.06 & 8.45$\pm$0.10 & 8.35$\pm$0.08 \\
3\dotfill & 3 & 1.3923 &  $10.06^{+0.06}_{-0.06}$ & -0.61$\pm$0.04 &  1.09$\pm$0.08 & 8.55$\pm$0.11 & 8.38$\pm$0.08 \\
4\dotfill & 3 & 1.3858 &  $10.63^{+0.07}_{-0.09}$ & -0.51$\pm$0.02 &       $>$ 0.62 & 8.61$\pm$0.10 & $<$ 8.53 \\
\enddata
\tablenotetext{a}{O3N2 $\equiv$ log$\{$([O III]$\lambda$5007/H$\beta$)/([N II] $\lambda$6584/H$\alpha$)$\}$.}
\tablenotetext{b}{Oxygen abundance deduced from the N2 relationship presented in \citet{pp04}.} 
\tablenotetext{c}{Oxygen abundance deduced from the O3N2 relationship presented in \citet{pp04}.}
\end{deluxetable*}
%---------------------------------------------------

%%%%%%%%---------------------------------------------------------------------------------------------

\section{The Offset in Diagnostic Line Ratios of High-Redshift Galaxies}\label{sec:offset}

\subsection{Emission-Line Diagnostics}

\begin{figure*}
  \centering
    \includegraphics[width=80mm]{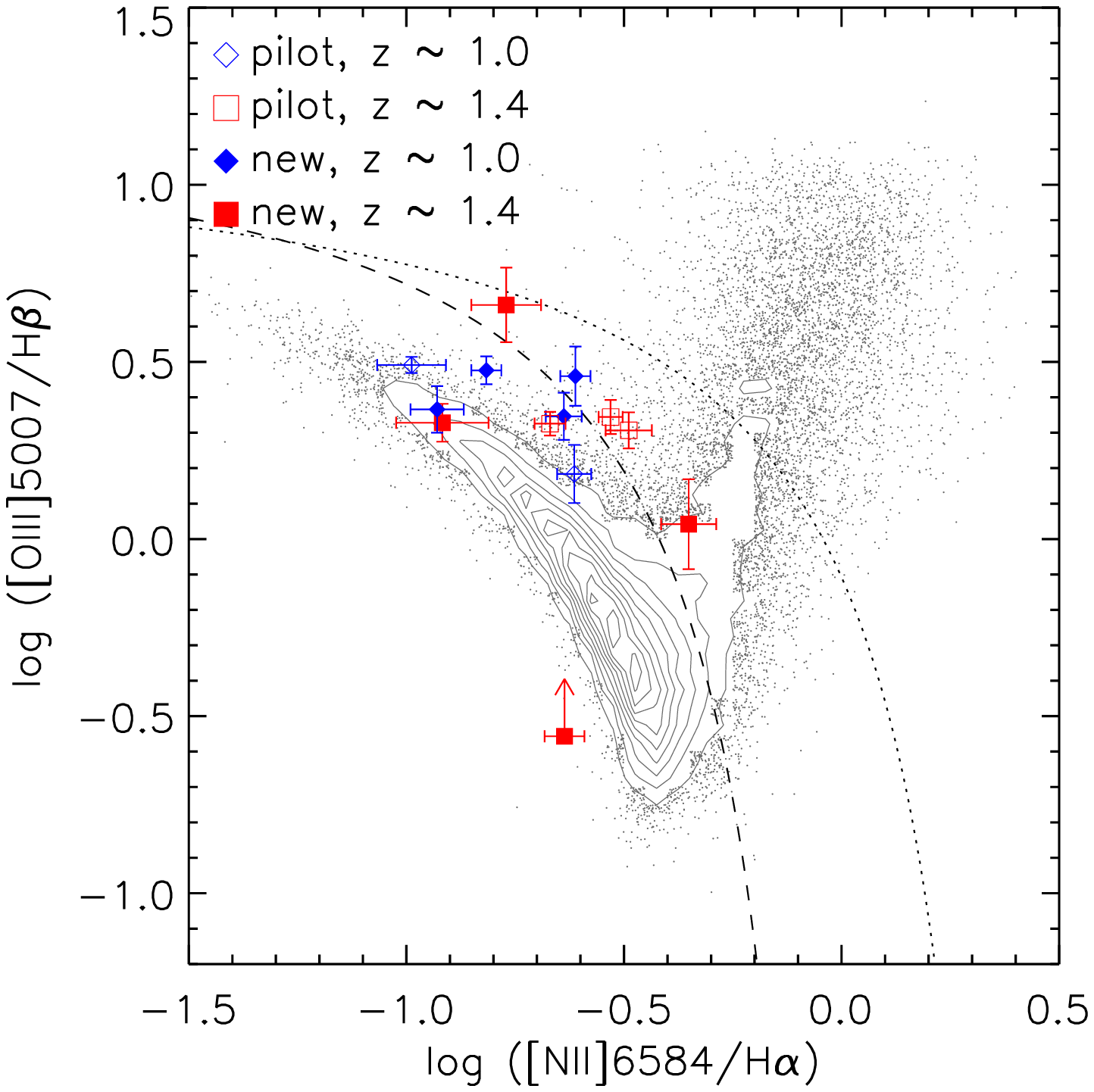}
    \includegraphics[width=80mm]{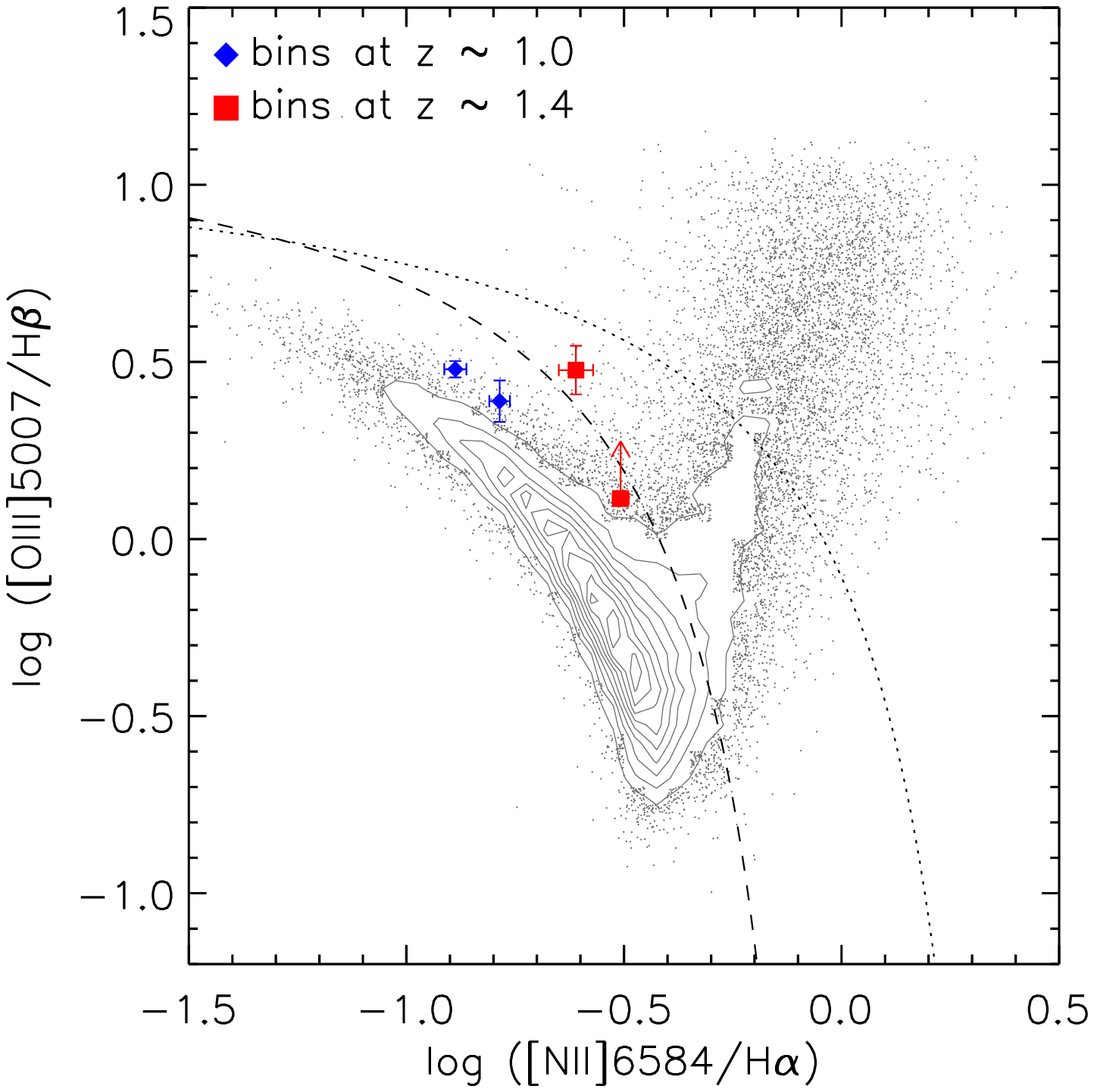}
    \caption{H II region log[N II]/H$\alpha$ vs. log[O III]/H$\beta$ diagram.
    On the left, the $z\sim$1.0-1.5 DEEP2 galaxies are shown as indicated on
    the plot. The right panel shows the average diagnostic line ratios from
    the composite spectra of DEEP2 objects. Note the bin with the highest
    [N II]/H$\alpha$ value only has a lower limit on [O III]/H$\beta$ because
    it includes the object 42021412. SDSS emission-line galaxies that satisfy
    the criteria described in \S \ref{subsec:sdsspreselect} are shown
    as grey contours and dots. The dashed line is an empirical demarcation
    from Ka03 based on the SDSS galaxies, whereas the dotted
    line is the theoretical limit for star-forming galaxies from Ke01.
    Nearby star-forming galaxies and H II regions form a well-defined excitation
    sequence of photoionization by massive stars, below and to the left of these
    curves. The $z \sim 1.0-1.5$ DEEP2 objects on average are offset from this
    excitation sequence, with objects at $z\sim$1.4 more offset than those at $z\sim$1.0.}
    \label{fig:subfig:bpt}
\end{figure*}

\begin{figure*}
  \centering
    \includegraphics[width=160mm]{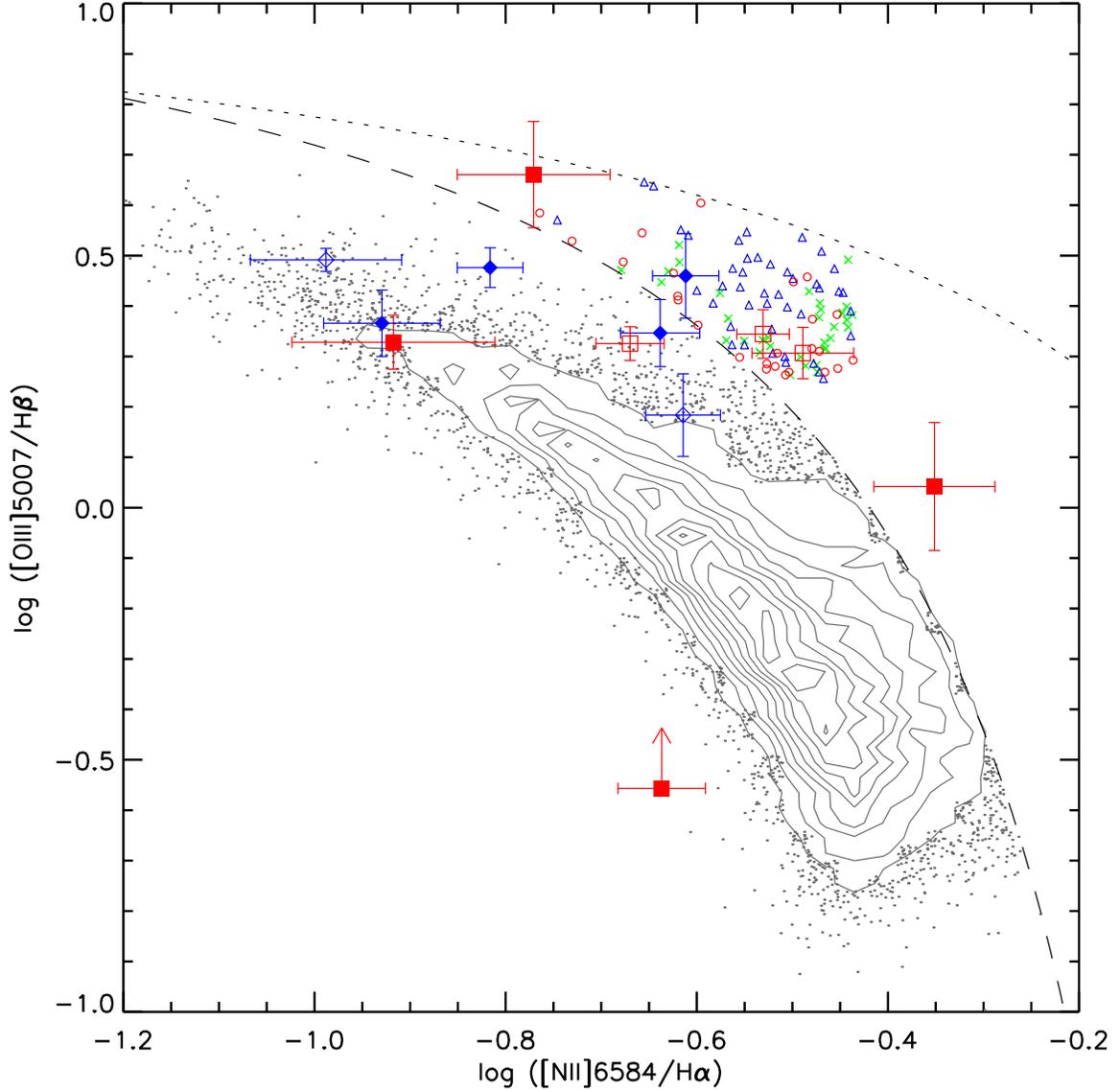}
    \caption{H II region diagnostic diagram: log[N II]/H$\alpha$ vs.
    log[O III]/H$\beta$. SDSS Main, Offset-AGN, Offset-ambiguous,
    and Offset-SF samples are shown as grey contours with dots, blue triangles,
    green crosses, and red circles, respectively (see \S \ref{subsec:sdsspreselect}
    for more information on sample selection rules). The $z\sim1.0-1.5$
    DEEP2 objects are also shown for a comparison using the same legends
    as in the left panel of Figure \ref{fig:subfig:bpt}, and the dotted
    and dashed lines have the same meanings as those in Figure \ref{fig:subfig:bpt}.}
    \label{fig:subfig:bptn2}
\end{figure*}

There is evidence that physical conditions in the H II regions of
high-redshift galaxies hosting intense star formation are
different from those of local H II regions (Paper I). The most
common method for probing H~II region physical conditions, and
discriminating between star-forming galaxies and AGNs, is based on
the positions of objects in the \citet*{bpt81} empirical
diagnostic diagrams (hereafter BPT diagrams). These plots feature
the optical line ratios [N II]/H$\alpha$, [O I]/H$\alpha$, [S
II]/H$\alpha$, and [O III]/H$\beta$, and have been updated by many
authors, including \citet{osterbrock85}, \citet{veilleux87},
\citet[][hereafter Ke01]{kewley01a}, and \citet[][hereafter
Ka03]{kauffmann03b}. A considerable fraction of high-redshift
star-forming galaxies at both z $\sim$ 1 (paper I) and z $\geq$ 2
\citep{erb06a} do not follow the local excitation sequence formed
by nearby H II regions and star-forming galaxies on the
emission-line diagnostic diagram of [N II]/H$\alpha$ versus [O
III]/H$\beta$; on average, they lie offset upward and to the
right. These differences must be taken into account when applying
empirically calibrated abundance indicators to galaxy samples at
different redshifts. Several possible causes for the offset have
been discussed in Paper I, including differences in the ionizing
spectrum, ionization parameter, electron density, and the effects
of shocks and AGNs.

In Figure \ref{fig:subfig:bpt}, [O III]/H$\beta$ and [N
II]/H$\alpha$ ratios are plotted in the left panel for the 13
galaxies in our sample with the full set of emission lines, and
the average [O III]/H$\beta$ and [N II]/H$\alpha$ ratios from the
N2+O3 composite spectra are also shown in the right panel for bins
at $z \sim 1.0$ and $1.4$, respectively. A subset of emission-line
objects from SDSS are also shown as grey contours and dots for
comparison. The dotted curve is from Ke01, representing a
theoretical upper limit on the location of star-forming galaxies
in the diagnostic diagram. The dashed curve is from Ka03 and
serves as an empirical discriminator between star-forming galaxies
and AGNs. On average, the H$\alpha$ flux of galaxies below this
curve should have $<1\%$ contribution from AGNs
\citep{brinchmann04}.
The effect observed in Paper I is still present in our larger
sample, in the sense that the $z \sim 1.0-1.4$ sample is, on
average, significantly offset from the excitation sequence formed
by star-forming galaxies from SDSS. Furthermore, the average
offset for the $z\sim 1.4$ objects is larger than for those at
$z\sim 1.0$. A similar, if not stronger, effect is observed in
star-forming galaxies at $z \sim 2$ \citep{erb06a}. As discussed
in Paper I, unaccounted-for stellar Balmer absorption is not the
explanation for the offsets in emission-line ratios, since the
corrections would shift the DEEP2 galaxies by no more than 0.1 dex
downward and by an insignificant amount in [N II]/H$\alpha$.

Isolating the causes of the offset of the high-redshift samples
from the local excitation sequence on the diagnostic diagram will
provide important insight into the physical conditions in distant
star-forming galaxies. These conditions also comprise an essential
systematic factor determining the emergent strong emission-line
ratios and associated calibration of chemical abundances.
Unfortunately, we do not currently have a large high-redshift
sample with available additional spectral features, stellar
population parameters, and morphological information, which is
needed for a direct study of how the diagnostic line ratios vary
as functions of galaxy properties. However, as seen in Figure
\ref{fig:subfig:bpt}, while only a tiny fraction of the SDSS
galaxies reside in the region of BPT parameter space inhabited by
our most extremely offset high-redshift galaxies, in between the
star-forming excitation sequence and the AGN branch, the sheer
size of the SDSS sample still results in a set of $\sim 100$ such
local objects. These objects can serve as possible local
counterparts for our DEEP2 objects, with the added benefit of high
S/N photometric, spectroscopic, and morphological information from
SDSS. We will make use of this detailed information to understand
the cause of the local galaxies' offset in the BPT diagram, and,
by extension, the likely cause of the offset among the
high-redshift galaxies.

In the following sections, we compare in detail these anomalous
SDSS objects against more typical SDSS star-forming galaxies.
Accordingly, we analyze possible causes for their offset in terms
of different physical conditions in the ionized regions, which
include H~II region electron density, hardness of the ionizing
spectrum, ionization parameter, the effects of shock excitation,
and contributions from an AGN. We further try to unravel possible
connections between physical conditions of these anomalous SDSS
objects and their host galaxy properties and use them to interpret
our observations of high-redshift star-forming galaxies.

\subsection{Local Counterparts: SDSS Main and Offset Samples}\label{subsec:sdsspreselect}

\begin{figure*}
  \centering
    \includegraphics[width=160mm]{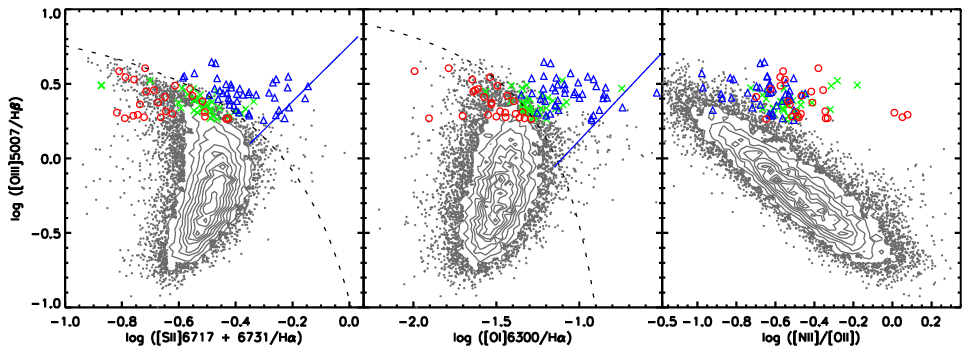}
    \caption{H II region diagnostic diagrams. Descriptions for SDSS objects
are the same as those in Figure \ref{fig:subfig:bptn2}. On the [O
I]/H$\alpha$ and [S II]/H$\alpha$ diagrams, the dotted (blue
solid) lines are empirical curves separating star-forming galaxies
(Seyferts) and AGN (LINERs) from \citet{kewley06}. The log[N
II]/[O II] vs. log[O III]/H$\beta$ diagram is useful in separating
ionization parameter and abundance but becomes insensitive to
ionization parameter above solar metallicity \citep{dopita00}. The
[N II]/[O II] flux ratio has been corrected for dust extinction,
given the wide spacing in wavelength between [N II] and [O II].}
    \label{fig:subfig:agnbpt}
\end{figure*}

Within this work we use the SDSS Data Release 4
\citep[DR4;][]{adelman06} spectroscopic galaxy sample, which
contains $u$-, $g$-, $r$-, $i$-, and $z$-band photometry and
spectroscopy of 567,486 objects. As described in
\citet{tremonti04}, emission-line fluxes of these galaxies were
measured from the stellar-continuum subtracted spectra with the
latest high spectral resolution population synthesis models by
\citet{bc03}.

Before being further divided into groups with different
emission-line diagnostic ratios, our sample was selected from the
567,486 galaxy DR4 sample according to the following criteria:

(1) Redshifts in the range $0.005 < z < 0.25$. (2) Signal-to-noise
ratio (S/N) $>$ 3 in the strong emission-lines [O II]
$\lambda\lambda$3726, 3729, H$\beta$, [O III]
$\lambda\lambda$4959, 5007, H$\alpha$, [N II] $\lambda$6584, and
[S II] $\lambda\lambda$6717, 6731; and S/N $>$ 1 in the weak
emission-line [O I] $\lambda$6300. In practice, $\sim 99$\% of the
objects also have [O I] $\lambda$6300 S/N$>$3. (3) The fiber
aperture covers at least 20\% of the total $g$-band photons. (4)
Stellar population parameter estimates are available in the
catalog derived using methods described by \citet{kauffmann03a}.

The first criterion on redshift range is the same as that adopted
by \citet{tremonti04}, which enables a fair comparison with the
mass-metallicity relation from their work. The second criterion
selects galaxies with well-measured emission-line fluxes, required
by our study based on emission-line diagnostic ratios. The S/N
values were obtained using errors of the emission-line fluxes,
which were first taken from the emission-line flux catalog on the
MPA DR4 Web site\footnote{See
http://www.mpa-garching.mpg.de/SDSS/} and then scaled by the
recommended factors (see the Web site for more details about the
emission-line flux error). The third criterion is required to
avoid significant aperture effects on the flux ratios
\citep{kewley05,tremonti04}. Lower aperture fraction can cause
significant discrepancies between aperture and global parameter
estimates. The fourth criterion enables a comparison of stellar
population properties with diagnostic emission-line ratios.
Roughly 30\% of the initial set of DR4 galaxies were ruled out
because they do not have stellar population parameter estimates;
an additional $\sim$20\% was cut out according to the redshift and
aperture-coverage constraints. Rules on the S/N in both strong and
weak emission lines removed another $\sim$45\% of the objects. In
all, these selection criteria leave us with $\sim$31,000
emission-line objects, shown as grey contours and dots in Figure
\ref{fig:subfig:bpt}, which is about 5\% of the 567,486 galaxy DR4
sample.

With the exception of one of the 13 DEEP2 objects, which lies on
the part of the [N II]/H$\alpha$ versus [O III]/H$\beta$ diagram
populated by many H II/AGN composites, the high-redshift galaxies
on average fall between the excitation sequence of typical SDSS
star-forming galaxies and the AGN branch. Although many of the
DEEP2 objects have a less extreme offset relative to typical SDSS
star-forming galaxies than those of the sample of \citet{erb06a},
and the most extreme $z\sim 1.4$ objects in Paper~I, there are
still several that are significantly offset from the main
sequence. We proceed by studying properties of the SDSS galaxies
with similar [N II]/H$\alpha$ and [O III]/H$\beta$ values to those
of the significantly offset DEEP2 objects, and comparing these
unusual objects with typical SDSS star-forming galaxies along the
excitation sequence. In this sense we further divide the SDSS
sample into the ``Main'', and ``Offset'' subgroups, as shown in
Figure \ref{fig:subfig:bptn2}.

Objects in the Main sample are selected to lie below the Ka03
empirical curve, whereas Offset objects are located in between the
Ka03 and the Ke01 curves. The Offset sample is also constrained to
have log([N II]/H$\alpha$) $\leq$ -0.44 and log([O III]/H$\beta$)
$\geq$ 0.26, according to the two emission-line ratios of the
DEEP2 object with the second largest [N II]/H$\alpha$ value. We do
not use the galaxy with the largest [N II]/H$\alpha$ value in our
DEEP2 sample for the selection criteria of the two line ratios,
because the number density in the composite region increases very
rapidly as the AGN branch is approached, yet most of our DEEP2
galaxies clearly do not fall in that regime. Also, it is worth
pointing out that, while they are significantly displaced from the
local excitation sequence, more than half of the DEEP2 objects
actually lie below the Ka03 curve. Indeed, the SDSS Offset sample
is constructed according to our DEEP2 objects that are offset the
most, in order to create a strong contrast with the Main sample.
It is also representative of the $z\sim 2$ galaxies observed by
\citet{erb06a}. As we describe in \S 5.6, the conclusions drawn
from this extreme sample are corroborated by the work of
\citet{brinchmann07}, where objects with less extreme offsets are
considered and therefore should be valid for typical objects in
our DEEP2 sample. Finally, the Offset sample only covers a certain
range of stellar masses, so we further select the Main control
sample according to the same stellar-mass range in order to ensure
a fair comparison. This leaves us with $\sim$21,000 objects for
the Main sample and 101 objects for the Offset sample.

Figure \ref{fig:subfig:agnbpt} displays how objects in the Offset
sample are distributed in the additional BPT diagrams featuring [O
I]/H$\alpha$ and [S II]/H$\alpha$. These plots indicate that
Offset objects selected solely on the basis of their [O
III]/H$\beta$ and [N II]/H$\alpha$ ratios span a diverse range of
properties in other physical parameter spaces. High [O
I]/H$\alpha$ and [S II]/H$\alpha$ both occur when there is a hard
ionizing radiation field, including significant contribution from
X-ray photons. In this case, there is an extended, partially
ionized zone, where H I and H II coexist, and [O I] and [S II] are
dominant forms of O and S. The extended zone of partially ionized
H does not exist in H II regions photoionized by OB stars
\citep{evans85,veilleux87}. High [O I]/H$\alpha$ and [S
II]/H$\alpha$ are also produced in gas that has been heated by
fast, radiative shocks, which also produce partially ionized
shock-precursor regions \citep{dopita95,dopita96}. Material in
supernova remnants provides an example of shocked gas. While it is
sensitive to shocks, [S II] is more susceptible to collisional
de-excitation than [O I], given its critical density
(2$\times10^3$ cm$^{-3}$), as opposed to that of [O I]
(2$\times10^6$ cm$^{-3}$), and therefore might be suppressed in
regions of high electron densities \citep{dopita97,kewley01b}.
Another point worth mentioning is that [O I]/H$\alpha$ should
reveal larger differences in regions ionized by hard spectrum as
opposed to that of stars, than [S II]/H$\alpha$ does. This is
because the ionization potential of [O I] matches that of H I, so
[O I] enhancement should reflect increased presence of partially
ionized zone. However, [S II] exists in both completely and
partially ionized zones. So the contrast between AGN- and
stellar-ionized regions for [S II]/H$\alpha$ is not as great as
that for [O I]/H$\alpha$.

In summary, the various locations of the Offset objects on the [O
I]/H$\alpha$ and [S II]/H$\alpha$ diagrams indicate different
levels of contribution from AGN and shock excitations to the
emerging spectra. We therefore further divide the Offset sample on
the [O I]/H$\alpha$ and [S II]/H$\alpha$ diagrams, according to
the theoretical scheme for classifying starburst galaxies and AGNs
\citep{kewley01a,kewley06}. It is worth noting here that all
objects in the Offset sample have S/N$>$3 in both [S II] and [O I]
emission lines, so this division should not be compromised by
spurious measurements. While there may still be a low-level
($<10$\%) AGN contribution to the spectra of objects classified as
starbursts by this scheme, it serves as a rough guide to the range
of properties in the Offset sample. This classification leaves us
with (1) ``Offset-AGN'' (43), where objects lie above both the
Ke01 curves of the two diagrams; (2) ``Offset-ambiguous'' (33),
where objects lie above one Ke01 curve and under the other one;
and (3) ``Offset-SF'' (25), where objects lie below both the Ke01
curves. The division of Offset-AGN and Offset-SF is only based on
hardness of the ionizing spectrum and the contribution of shock
excitation. We show in Section \ref{sec:sub:comparison} that these
two offset samples have different host galaxy properties.

\subsection{H II Region Emission-Line Diagnostic Ratio, Physical Conditions, and Galaxy Properties}

\begin{figure*}
  \centering
    \includegraphics[width=160mm]{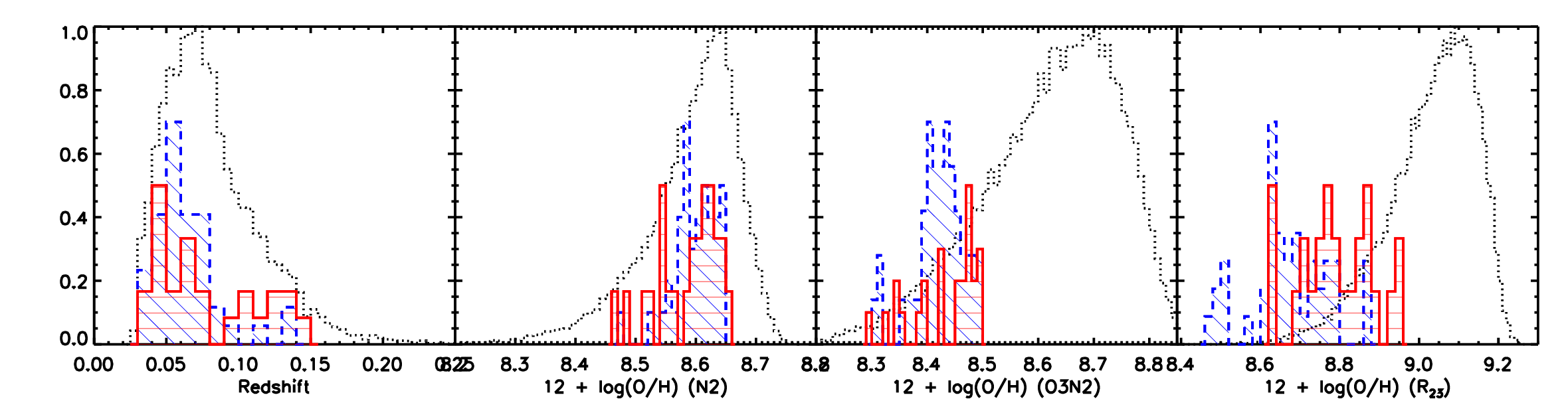}
    \includegraphics[width=160mm]{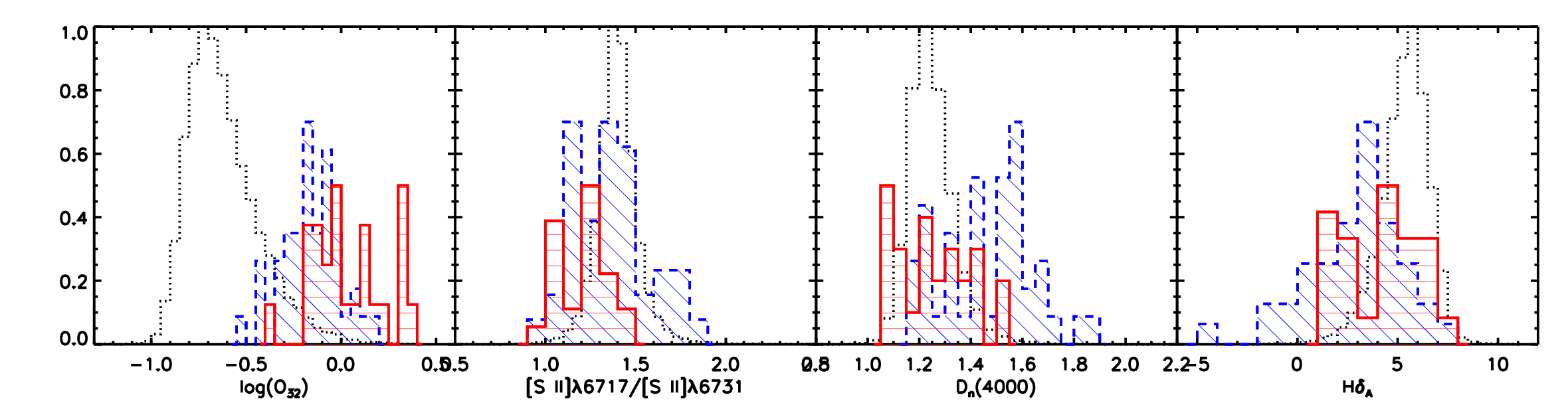}
    \includegraphics[width=160mm]{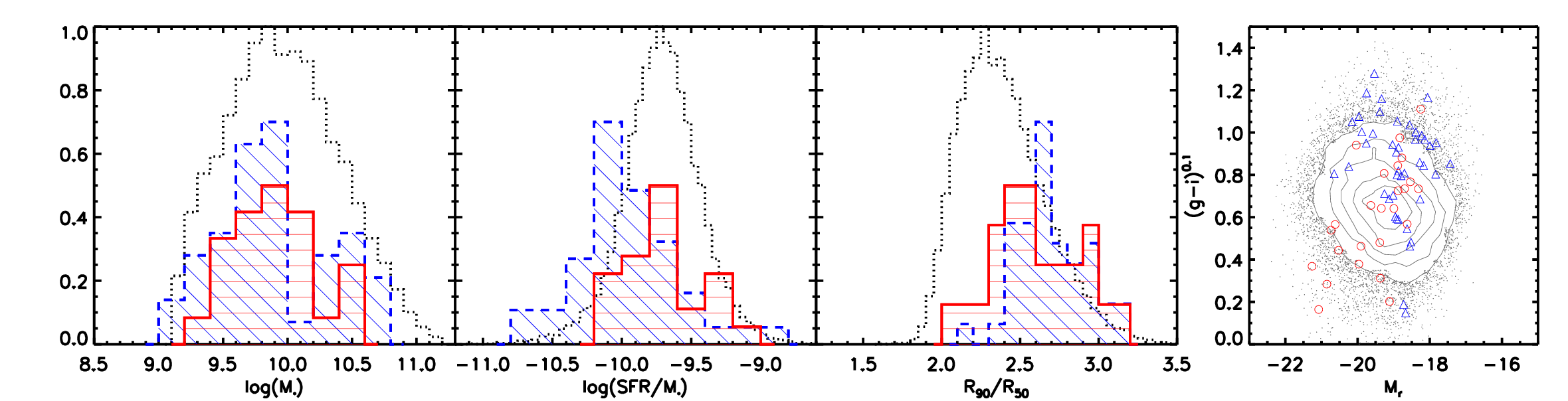}
    \includegraphics[width=160mm]{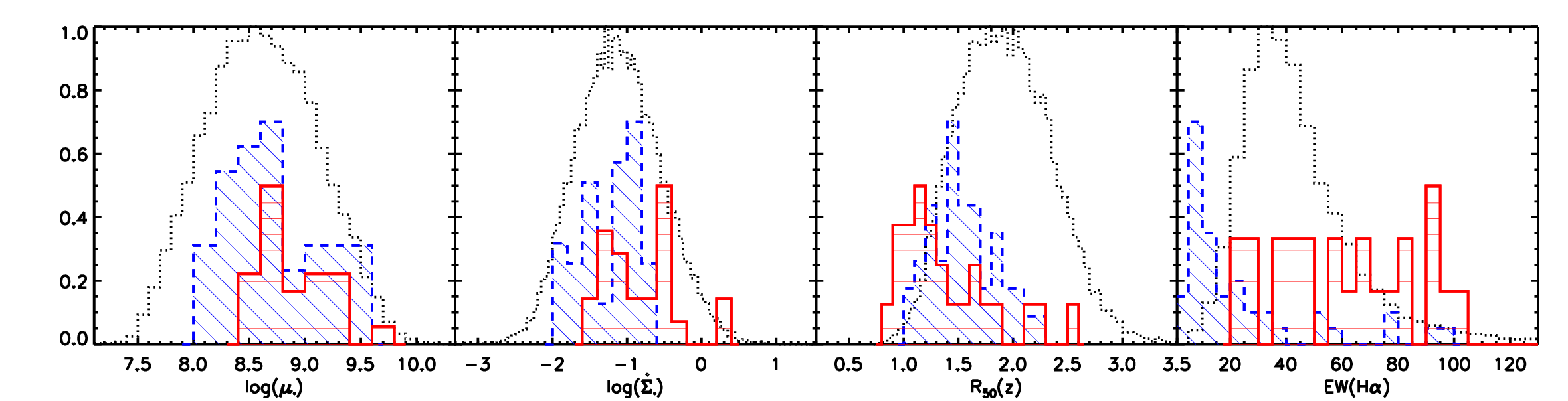}
    \caption{Distribution of row (1): redshift and metallicities
    based on various strong-line indicators; row (2): ionization-parameter
    indicator log($O_{32}$), electron density indicator [S II] $\lambda6717$/[S II] $\lambda6731$,
    D$_n(4000)$ and H$\delta_{A}$; row (3): stellar mass,
    fiber-corrected specific star formation rate, $r$-band Petrosian
    concentration index $C = R_{90}/R_{50}$, and color-magnitude diagram; row
    (4): surface mass density log($\mu_{\star}$), SFR surface density
    log($\dot{\Sigma}_{\star}$), $R_{50}(z)$, which is the radius enclosing
    50\% of the Petrosian $z$-band luminosity of a galaxy, and H$\alpha$
    emission equivalent widths, for the SDSS Main ({\it black dotted line}),
    Offset-AGN ({\it blue dashed line}), and Offset-SF samples ({\it red solid line}; see \S \ref{subsec:sdsspreselect}
    for more information). For clarity, the distribution functions are
    normalized to three different levels, with the maxima being 1.0,
    0.7, and 0.5 for the SDSS Main, Offset-AGN, and Offset-SF samples, respectively.}
    \label{fig:subfig:dis}
\end{figure*}

In order to determine the origin of the observed difference in
emission-line ratios between the Main and Offset samples, we
consider a large set of galaxy properties. Measurements of
emission lines as well as photometric, physical, and environmental
properties of stellar population are available for our SDSS
samples. In this section, we describe the galaxy properties of
interest, while in the following section, a comparison of the Main
and Offset sample distributions in these properties is presented.

First, using emission-line ratios, we examine the main factors
controlling the emission-line spectrum in an H~II region, i.e.,
the gas phase metallicity, the shape or hardness of the ionizing
radiation spectrum, and the geometrical distribution of gas with
respect to the ionizing sources, which can be represented by the
mean ionization parameter and electron density \citep{dopita00}.
Several H~II region emission-line ratios can be used as
diagnostics of these factors.

Apart from the N2 and O3N2 indicators, we also use the ratio N2O2
$\equiv$ log([N II] $\lambda$6584/[O II] $\lambda\lambda$3726,
3729), suggested by \citet{vanzee98}, as an abundance diagnostic
for SDSS objects, because N2O2 is virtually independent of
ionization parameter and also strongly sensitive to metallicity.
This indicator is monotonic between 0.1 and over 3.0 times solar
metallicity \citep{dopita00,kewley02}. Concerns about reddening
correction and reliable calibration over such a large wavelength
baseline have hampered the use of this ratio. As shown in
\citet{kewley02}, however, the use of classical reddening curves
and standard calibration are quite sufficient to allow this [N
II]/[O II] diagnostic to be used as a reliable abundance
indicator. The \citet{bresolin07} calibration of N2O2 is used to
infer oxygen abundance. We also use the $R_{23} \equiv$ log\{([O
II] $\lambda$3727 + [O III] $\lambda\lambda$4959, 5007)/H$\beta$\}
parameter, introduced by \citet{pagel79}, as an abundance
indicator, to make comparison with several works
\citep{lilly03,savaglio05}, although this indicator has some
well-documented drawbacks \citep[e.g.][]{kobulnicky99,kewley02}.

$O_{32} \equiv$ log([O III] $\lambda$5007/[O II]
$\lambda\lambda$3726, 3729) is used as an indicator of the
ionization parameter \citep{mcgaugh91,dopita00}. However, $O_{32}$
is sensitive to both ionization parameter and abundance; both low
ionization parameter and high abundances produce low values of
$O_{32}$ \citep{dopita00,kewley02}. Thus, when $O_{32}$ is used as
a diagnostic of the relative ionization parameters of different
samples, the metallicities of these samples must also be taken
into account for a fair comparison. A diagnostic plot, for
example, N2O2 versus $O_{32}$, can be used to separate ionization
parameter and abundance. The ratio [S III]/[S II] provides another
diagnostic of the ionization parameter, which is independent of
metallicity except at very high values of the ratio
\citep{kewley02}. Unfortunately, the SDSS spectra do not cover [S
III] $\lambda\lambda$9069, 9532 from SDSS, so it is not included
here.

Finally, [S II] $\lambda6717$/[S II] $\lambda6731$ is adopted as
an electron-density indicator. We do not use the density-sensitive
ratio, [O II] $\lambda3726$/[O II] $\lambda3729$, since this
doublet is typically blended in SDSS spectra.

We note that emission-line ratios, such as $O_{32}$, N2O2, and
$R_{23}$ have large separations in wavelength and thus need to be
corrected for dust reddening. The Balmer decrement method with the
reddening curve of \citet{calzetti00} is used to correct these
quantities for dust extinction, assuming an intrinsic ratio of
$H\alpha/H\beta=2.78$, appropriate for $T_e=10,000$ K.

Next, we turn to the question of the stellar populations and
structural and environmental properties of the Main and the Offset
sample host galaxies. To examine stellar ages, we adopt the narrow
definition of the 4000 ${\rm \AA}$ break denoted as D$_n(4000)$
\citep{balogh99}, which is small for young stellar populations and
large for old, metal-rich galaxies \citep{kauffmann03a}. We use
the H$\delta_{A}$ index as an indicator of recent starburst
activities \citep[e.g.][]{kauffmann03a,worthey97}. Strong
H$\delta$ absorption lines arise in galaxies that experienced a
burst of star formation that ended $\sim$0.1-1 Gyr ago. For galaxy
morphology, we use the standard concentration parameter defined as
the ratio $C = R_{90}(r)/R_{50}(r)$, where $R_{90}(r)$ and
$R_{50}(r)$ are the radii enclosing 90\% and 50\% of the Petrosian
$r$-band luminosity of the galaxy. For galaxy size, we study
$R_{50}(z)$, the radius enclosing 50\% of the Petrosian $z$-band
luminosity of a galaxy. We also examine the surface mass density
$\mu_{\star}$, defined as $M_{\star}/[2\pi R^2_{50}(z)]$
\citep{kauffmann03c}, the SFR surface density
$\dot{\Sigma}_{\star}\equiv {\rm SFR}/[2\pi R^2_{50}(r)]$, where
we use a radius defined in the $r$ band rather than the $z$ band,
as it is more appropriate for H$\alpha$ luminosities. Additional
parameters include the fiber-corrected specific star formation
rate SFR/M$_{\star}$, H$\alpha$ equivalent width (EW), and the
color-magnitude diagram $(g-i)^{0.1}$ versus $M_r$, where
$(g-i)^{0.1}$ denotes the $(g-i)$ color $k$-corrected to $z =
0.1$, and $M_r$ stands for the $k$-corrected $r$-band absolute
magnitude \citep{kauffmann03a}. In order to search for any
environmental dependence of the Offset sample properties, we count
the number of spectroscopically observed galaxies that are located
within 2 Mpc in projected radius and $\pm$500 km s$^{-1}$ in
velocity difference from each object in our SDSS samples. Here we
follow the procedures described in \citet{kauffmann04},
constructing a volume-limited tracer sample using the SDSS DR4
data.

\subsection{Comparison with Typical SDSS Star-forming Galaxies}\label{sec:sub:comparison}

\begin{figure*}
\epsscale{1.} \plotone{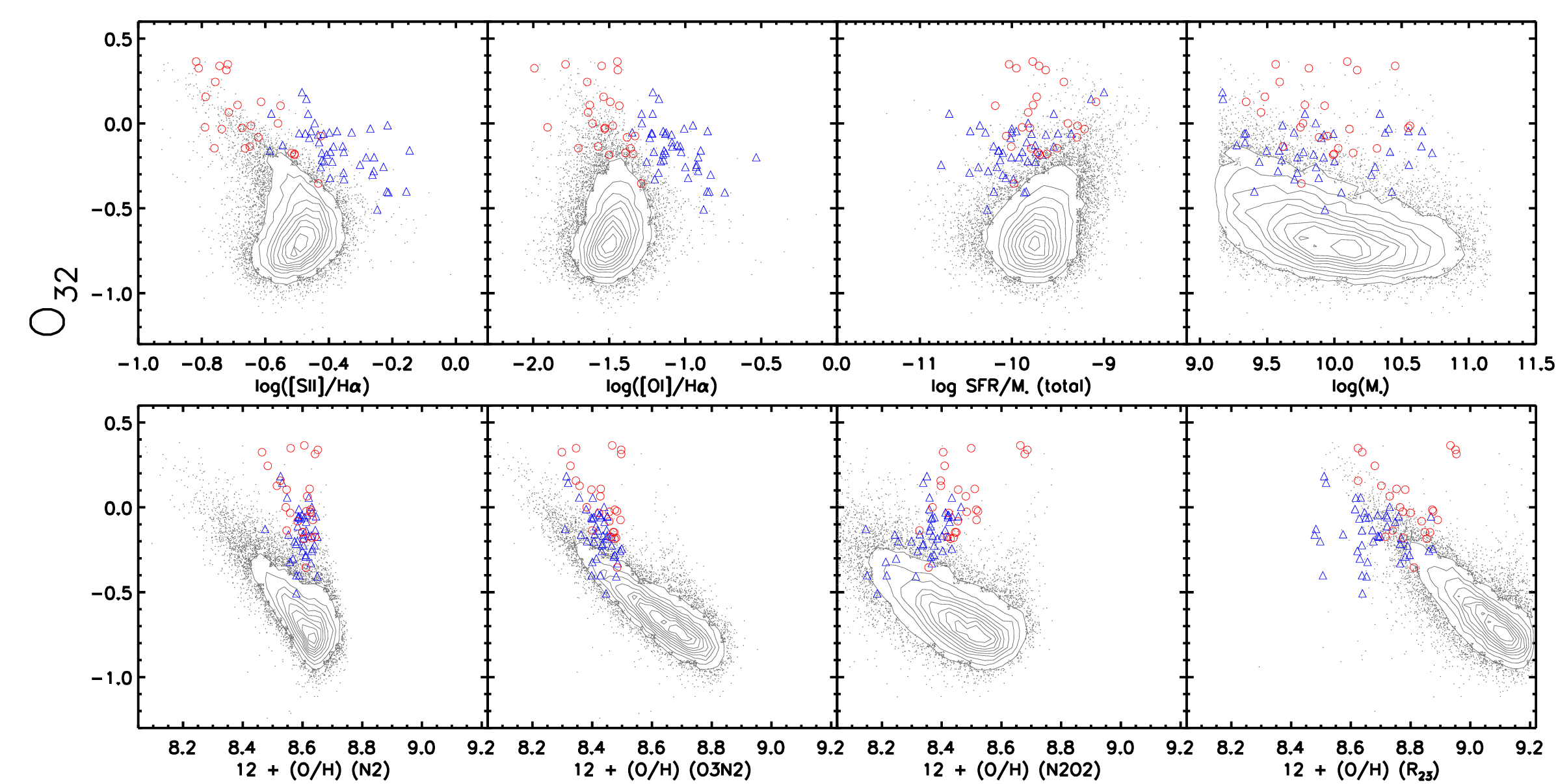} \caption{Ionization-parameter
indicator $O_{32}$, as a function of various parameters.
Descriptions are the same as those in Figure
\ref{fig:subfig:agnbpt}.\label{o32_dep}}
\end{figure*}

In this section, we compare the properties of Offset and Main
sample objects, in terms of their H~II region physical conditions,
stellar populations, structural parameters, and environments. In
particular, we find striking differences in H~II region ionization
parameter, electron density, galaxy size, and star formation rate
surface density. Figure \ref{fig:subfig:dis} shows relative
distributions of diagnostic-line ratios and galaxy properties of
the SDSS Main, Offset-AGN, and Offset-SF samples. For clarity, the
distribution functions are normalized to three different levels,
with the maxima being 1.0, 0.7 and 0.5 for the SDSS Main,
Offset-AGN, and Offset-SF samples, respectively.

First, we consider H~II region physical conditions. Compared to
the Main sample, both Offset samples on average have larger
ionization parameters. The Offset-SF and Offset-AGN samples have
$O_{32} = 0.0$ and -0.2, respectively, as opposed to -0.7 for the
Main sample. Here and throughout, we refer to the median values of
the distributions. Since $O_{32}$ also depends on metallicity, we
need to make a fair comparison for the ionization parameter at a
fixed metallicity. Figure \ref{o32_dep} shows $O_{32}$ as a
function of several emission-line diagnostic ratios and galaxy
properties. In the lower four panels, we can see that, regardless
of which metallicity indicator is used, the Offset-SF sample has
bigger $O_{32}$ values than the Main sample at a fixed strong-line
indicator value. Therefore, the Offset-SF sample has larger
average ionization parameter than Main sample objects having
similar metallicities, independent of the metallicity indicator.
Compared with the Main sample, all indicators except $R_{23}$
suggest that the Offset-AGN has higher ionization parameters at
fixed metallicities.

The Offset-SF sample also has significantly larger electron
densities, on average, than the Main sample ([S II]
$\lambda6717$/[S II] $\lambda6731$ of $\sim$1.23 compared to
$\sim$1.40, corresponding to electron densities of $\sim$208
cm$^{-3}$ compared to $\sim$47 cm$^{-3}$), whereas the Offset-AGN
sample has only slightly higher electron densities than those of
the Main sample ([S II] $\lambda6717$/[S II] $\lambda6731$ of
$\sim$1.37, corresponding to electron densities of $\sim$67
cm$^{-3}$). As discussed in Paper I, for fixed ionization
parameter, metallicity, and input ionizing spectrum, the
photoionization models presented in \citet{kewley01a} display a
dependence on electron density, in the sense that model grids with
higher electron density have an upper envelope in the space of [O
III]/H$\beta$ versus [N II]/H$\alpha$ that is offset upward and to
the right, relative to model grids with lower electron density.
This theoretical shift in the BPT diagram due to increased
electron density is qualitatively reflected in the properties of
Offset-SF objects.

Next, we consider photometric and spectroscopic stellar population
properties. The Offset-SF sample is similar to the Main sample in
terms of median color [$(g-i)^{0.1}$ $\sim$ 0.64 compared to
$\sim$0.66], stellar age [the same value of D$_n$(4000) = 1.24],
and fiber-corrected specific star formation rates (the same value
of $\log((SFR/M_{\star})\mbox{ yr}^{-1}) = -9.7$), whereas the
Offset-AGN sample has redder colors [$(g-i)^{0.1}$ of $\sim$0.86],
older stellar ages [D$_n$(4000) = 1.50], and lower specific star
formation rates ($\log((SFR/M_{\star})\mbox{ yr}^{-1}) = -10.0$).
Also, the Offset-SF sample has a larger H$\alpha$ EW than the Main
sample (62 ${\rm \AA}$ compared to 40 ${\rm \AA}$), whereas the
Offset-AGN sample has a much smaller H$\alpha$ EW (15${\rm \AA}$).
The Offset-SF sample has a smaller burst fraction in stellar mass
than the Main sample (H$\delta_{A}$ of 4.3 compared to 5.5), while
the discrepancy between the Offset-AGN and Main samples is even
larger (H$\delta_{A}$ of 3.3 compared to 5.5).

In terms of galaxy structure, the Offset-SF sample has larger
concentration ($C$ = 2.58 compared to 2.36), smaller half-light
radii [$R_{50}(z)$ of 1.2 kpc compared to 1.9 kpc], higher surface
stellar mass density [log($\mu_{\star}$) of 8.78 M$_{\odot}$
kpc$^{-2}$ compared to 8.61 M$_{\odot}$ kpc$^{-2}$], and higher
SFR surface density [log($\dot{\Sigma}_{\star}$) of -0.86
M$_{\odot}$ yr$^{-1}$ kpc$^{-2}$ compared to -1.12 M$_{\odot}$
yr$^{-1}$ kpc$^{-2}$] than the Main sample, whereas the Offset-AGN
has larger concentration ($C$ of 2.66), moderately smaller sizes
[$R_{50}(z)$ of 1.5 kpc], moderately higher surface stellar mass
density [log($\mu_{\star}$) of 8.68 M$_{\odot}$ kpc$^{-2}$], and
similar SFR surface density [log($\dot{\Sigma}_{\star}$) of -1.12
M$_{\odot}$ yr$^{-1}$ kpc$^{-2}$].

In terms of environment, we find that 90\%-95\% of objects in the
Main, Offset-SF, and Offset-AGN samples have zero or one neighbor,
the lowest density bin in \citet{kauffmann04}. Furthermore, the
distributions of environments for the Offset samples are similar
to that of the Main sample. Therefore, it appears that the vast
majority of galaxies considered here reside in the lowest density
environment defined by \citet{kauffmann04}. This result is not
surprising, as our SDSS samples all contain emission-line
galaxies, which are more likely to be found in low-density
environments \citep{kauffmann04}. Currently we have too few Offset
objects to draw any solid conclusion about the question on galaxy
environmental dependence.

\citet{groves06} suggest that the offset on the BPT diagram seen
in some high-redshift galaxies is mostly caused by contribution
from AGNs, as they found properties of their candidate
low-metallicity H II/AGN composites similar to those of the host
galaxies of AGNs \citep{kauffmann03b,heckman04}; these objects
have on average significantly higher stellar masses, older stellar
populations, and redder colors than the sample of pure
star-forming galaxies used for comparison. \citet{groves06} arrive
at this conclusion after constructing their offset sample in a
region bounded by the Ka03 and the Ke01 curves in the [N
II]/H$\alpha$ versus [O III]/H$\beta$ diagram, and the Seyfert
branch [log([O III]/H$\beta$) $\geq$ 3log([N II]/H$\alpha$)]. When
we reproduced their offset sample, we found the median values of
log([N II]/H$\alpha$) $\sim$ -0.286 and log([O III]/H$\beta$)
$\sim$ -0.024, whereas the corresponding values for our DEEP2
objects are -0.638 and 0.344. As the density of objects increases
rapidly towards increasing [N II]/H$\alpha$, the offset sample
definition of \citet{groves06} is weighted heavily towards H
II/AGN composites and not necessarily representative of the
average [N II]/H$\alpha$ or [O III]/H$\beta$ of high-redshift
galaxies. Furthermore, it is not clear that their ``typical''
comparison sample, which was defined as all galaxies within
$\pm$0.05 dex of [-0.55, 0.10] in the [N II]/H$\alpha$ versus [O
III]/H$\beta$ diagram, represents a fair one, as they have not
controlled for any galaxy property. Given the strong correlations
among galaxy properties, it is crucial to compare galaxies that
are similar in at least some basic parameters.
Here we want to emphasize that in order to make a controlled
comparison, the Main sample was selected to have similar range and
median of stellar masses as that of the Offset sample. The Main
sample also has a similar median [N II]/H$\alpha$ value. With this
controlled comparison of typical and Offset SDSS objects, we find
that the unusual objects on the BPT diagram are populated not only
by likely AGN hosts (Offset-AGN) but also by the Offset-SF sample,
most of which do not resemble typical AGN-host galaxies.
Furthermore, there is no segregation of the two offset samples in
the [N II]/H$\alpha$ versus [O III]/H$\beta$ diagram.

As suggested in \citet{groves06}, another possible test of the
presence of an AGN would require the detection of either He II
$\lambda$4686 or [Ne V] $\lambda$3426 lines. The [Ne V]
$\lambda$3426 line is not redshifted into the SDSS band for the
majority of our objects: we have only four such objects in
Offset-AGN, of which the spectra near [Ne V] $\lambda$3426 are too
noisy to provide any meaningful constraints. The He II
$\lambda$4686 line is expected to be very weak: a 20\% AGN
contribution to H$\beta$ implies He II/H$\beta$ = 0.05. Note that
besides the AGN photoionization, possible mechanisms for producing
He II emission also include hot stellar ionizing continua and
shock excitation \citep{garnett91}. Another important goal is to
infer metallicities of the Offset-SF objects using a method that
is independent of strong-line indicators, since these objects have
abnormal strong-line ratios. A more calibration-independent way to
address metallicity is through the classic $T_e$ method, which
relies on measuring weak auroral lines. Both the test of the
presence of AGNs, and the determination of metallicity with the
direct $T_e$ method, require the measurement of very weak lines.
These measurements become feasible through the use of composite
spectra.

\subsection{SDSS Composite Spectra}

\begin{figure*}
  \centering
    \includegraphics[width=150mm]{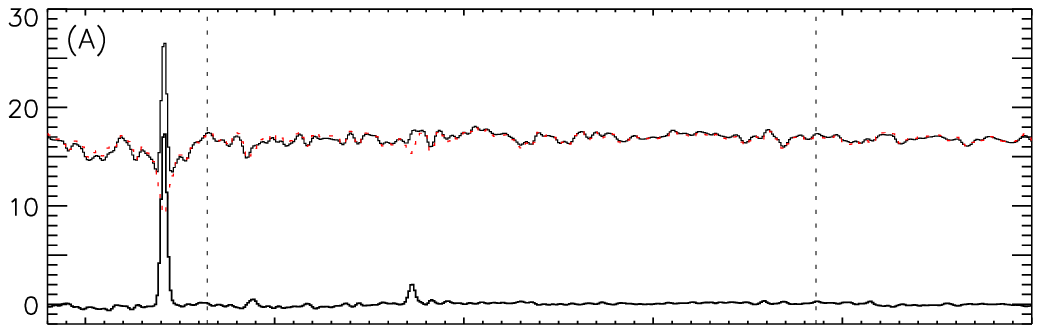}
    \includegraphics[width=150mm]{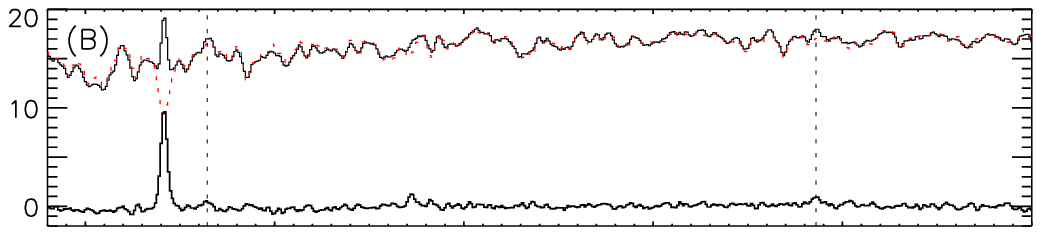}
    \includegraphics[width=150mm]{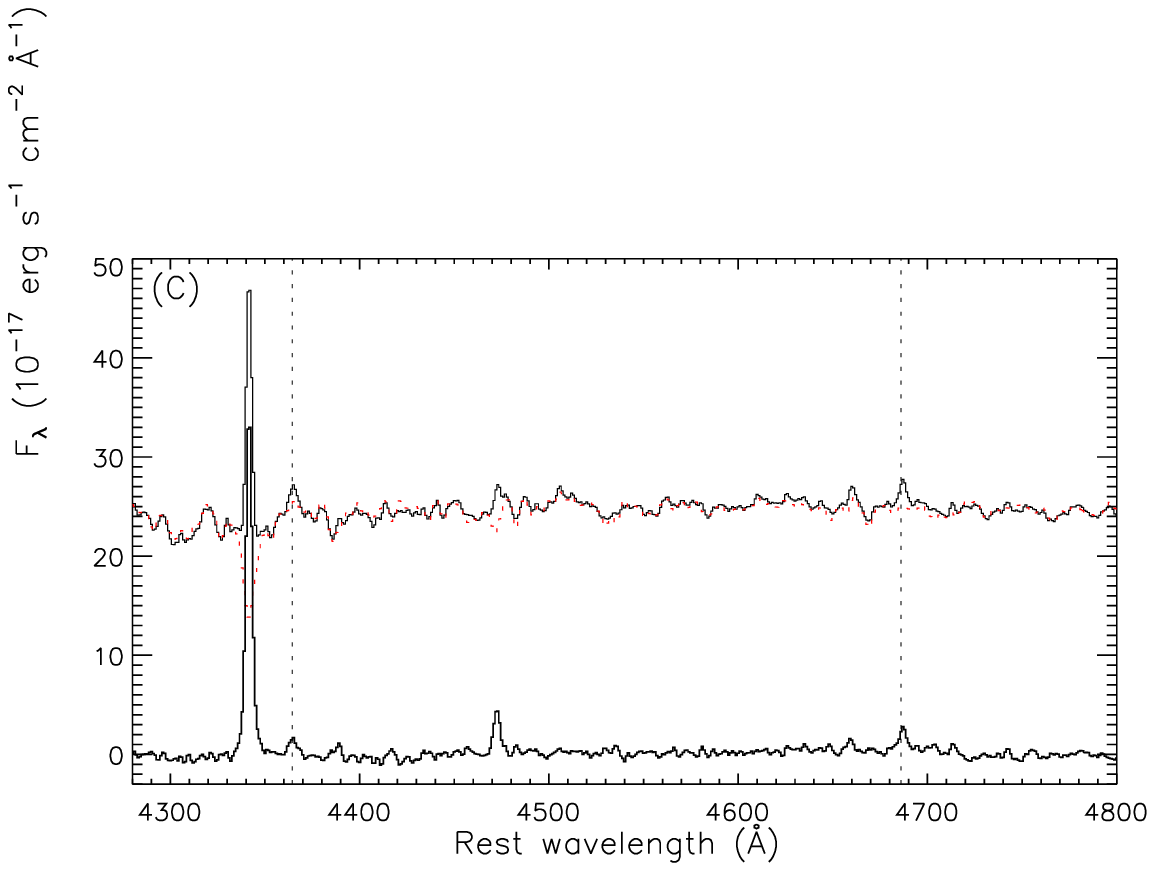}
    \caption{SDSS composite spectra. For clarity, only the portion
    of the spectrum containing the weak emission lines of interest
    is presented here. The original composite and stellar-continuum-subtracted
    spectra for samples (A) Main, (B) Offset-AGN, and (C)
    Offset-SF
    are shown as black solid curves. Stellar-continuum fits using
    \citet{bc03} population synthesis models are plotted as red dotted
    curves. The [O III] $\lambda$4363 and He II $\lambda$4686 lines are marked by dotted lines.}
    \label{fig:sdsscompspec}
\end{figure*}

In order to measure He II $\lambda$4686 as another test of
possible AGN contribution to the ionizing spectrum, and [O III]
$\lambda$4363 to infer the oxygen abundance using the $T_e$
method, we construct composite spectra for both Offset-SF and
Offset-AGN objects. For comparison, we also make composite spectra
for the Main sample. We adopt the same method of making composite
spectra used for our DEEP2 objects, employing median scaling to
preserve the relative fluxes of the emission features. The
variance spectrum is combined in the same way and the error for
the corresponding composite spectrum is the square root of the
composite variance spectrum divided by the number of objects.

The stellar continuum was modelled over the rest-frame wavelength
range of 3800 - 7500 ${\rm \AA}$ and then subtracted from the
composite spectra, with a method similar to the one described in
\citet{tremonti04} and \citet{brinchmann04}. This technique
\citep[][private communication]{charlot07} relies on the
\citet{bc03} stellar population synthesis model, but with two
major improvements. First, the number of the metallicity grids was
expanded from three to four; second, and most importantly, the
stellar continuum fitting was carried out using both the STELIB
\citep{leborgne03} and the MILES \citep{sanchez06} libraries of
observed stellar spectra. Models using the MILES library fit the
data better than those with the STELIB one, in terms of the
agreement around Balmer lines. It is especially essential to
obtain a reasonable stellar continuum fit near H$\gamma$, which is
very close in wavelength to the weak [O III] $\lambda$4363 auroral
line. Models using the STELIB library tend to over-estimate
stellar continua, especially around Balmer lines. We therefore
adopt the stellar continuum models based on the MILES library.

The resultant continuum-subtracted spectra, along with the
original composite spectra and the best-fit model of the stellar
continua, are shown in Figure (\ref{fig:sdsscompspec}). We only
present portions of the entire composite spectra to enable
scrutiny of the relevant weak lines ([O III] $\lambda$4363 and He
II $\lambda$4686) and our capability of making a reasonably good
stellar continuum fit. The corresponding error spectra are
estimated by combining the measurement error for the composite
spectra and the systematic uncertainties from the stellar
continuum fitting. The systematic error from the stellar continuum
fitting is estimated using the rms of a portion of the
continuum-subtracted spectrum, which is relatively free of
emission-line features. This method of estimating continuum fit
error is motivated by the fact that the rms would be zero if the
fitting was ideally good and if there were no emission-line
features. At all of the wavelengths considered, the systematic
error is significantly larger than the measurement error for the
composite spectra and therefore dominates the total error budget.
Emission-line fluxes and flux ratios with the total uncertainties
are given in Table \ref{tab:sdsscomp}.

There are a few caveats that must be mentioned in the analysis of
composite spectra described here. First, we are averaging over
tens of thousands of galaxies for the SDSS Main sample and dozens
of galaxies for the Offset-SF and Offset-AGN samples; second, for
each single galaxy, we are dealing with integrated spectra
containing contributions from multiple H II regions, which may
show a range of metallicities, ionization parameters, ionizing
spectra, and electron densities. Also, even though we select the
samples to have fiber apertures covering more than 20\% of the
total $g$-band photons, because of the presence of radial
gradients in galaxy properties, uncertainties still remain since
the spectrum is weighted towards the nucleus. Despite these
caveats, however, we want to emphasize that the analysis of
composite integrated fibre-based spectra is still meaningful in
terms of determining average properties and the relative
differences among samples.

%------------------------------------------
\begin{deluxetable*}{lccccccc}
\tabletypesize{\scriptsize}
\tablecaption{Emission-line flux measurements of SDSS composite spectra.\label{tab:sdsscomp}}
\tablewidth{0pt}
\tablehead
{
\colhead{~~~~~~~~~Sample~~~~~~~~~} & 
\colhead{F$_{{\rm [O II]}\lambda\lambda3726, 3729}$\tablenotemark{a}} & 
\colhead{F$_{{\rm [O III]}\lambda4363}$\tablenotemark{a}} & 
\colhead{F$_{{\rm He II}\lambda4686}$\tablenotemark{a}} & 
\colhead{F$_{{\rm H}\beta}$\tablenotemark{a}} &
\colhead{F$_{{\rm [O III]}\lambda4959}$\tablenotemark{a}} & 
\colhead{F$_{{\rm [O III]}\lambda5007}$\tablenotemark{a}} &
\colhead{F$_{{\rm [O I]}\lambda6300}$\tablenotemark{a}}
}
\startdata
Main\dotfill       & 327.3$\pm$0.3 & 0.6$\pm$0.2 &  1.4$\pm$0.2 & 153.5$\pm$0.2 &  45.2$\pm$0.1 & 137.0$\pm$0.3 & 19.1$\pm$0.2 \\
Offset-AGN\dotfill & 271.9$\pm$0.8 & 2.8$\pm$0.5 &  4.9$\pm$0.5 &  90.4$\pm$0.5 &  86.4$\pm$0.2 & 261.7$\pm$0.6 & 24.3$\pm$0.6 \\
Offset-SF\dotfill  & 629.1$\pm$1.0 & 7.1$\pm$0.6 & 12.5$\pm$0.7 & 321.0$\pm$0.7 & 266.6$\pm$0.3 & 807.4$\pm$0.8 & 38.8$\pm$0.7 \\
\hline
~~~~~~~~~Sample~~~~~~~~~ &  F$_{{\rm H}\alpha}$\tablenotemark{a} & F$_{{\rm [S II]}\lambda6717}$\tablenotemark{a} & F$_{{\rm [S II]}\lambda6731}$\tablenotemark{a} & $\frac{{\rm F_{[O III] \lambda \lambda 4959, 5007}}}{{\rm F_{[O III] \lambda 4363}}}$\tablenotemark{b}  & $\frac{{\rm F_{He II \lambda 4686}}}{{\rm F_{H\beta}}}$ & $\frac{{\rm F_{[O I] \lambda 6300}}}{{\rm F_{H\alpha}}}$ & $\frac{{\rm F_{[S II]\lambda\lambda6717+6731}}}{{\rm F_{H\alpha}}}$ \\
\hline
          &   &    &    &   &     &   &    \\
Main\dotfill         &  633.6$\pm$0.3 & 117.6$\pm$0.3  & 83.9$\pm$0.3  & 249$\pm$82 &  0.009$\pm$0.001  & 0.0301$\pm$0.0004 & 0.3180$\pm$0.0006  \\
Offset-AGN\dotfill   &  337.3$\pm$0.7 &  80.6$\pm$0.6  & 58.5$\pm$0.6  & 110$\pm$19 &  0.054$\pm$0.006  & 0.0720$\pm$0.0018 & 0.4124$\pm$0.0028   \\
Offset-SF\dotfill    & 1272.8$\pm$0.9 & 170.6$\pm$0.8  & 137.3$\pm$0.8 & 129$\pm$11 &  0.039$\pm$0.002  & 0.0305$\pm$0.0006 & 0.2419$\pm$0.0009  \\
\enddata
\tablenotetext{a}{Emission-line flux and 1 $\sigma$ error in units of $10^{-17}$ ergs s$^{-1}$ cm$^{-2}$ from SDSS composite spectra.}
\tablenotetext{b}{Dust-reddening corrected using Balmer decrement method.}
\end{deluxetable*}
%---------------------------------------------------

\subsubsection{AGN Contribution}

The composite spectra we construct can be used to address the
level at which AGN ionization contributes to the anomalous
emission-line ratios, discovered in both nearby SDSS Offset
objects and high-redshift DEEP2 galaxies. Based on their [S
II]/H$\alpha$ and [O I]/H$\alpha$ ratios, we expect the Offset-AGN
objects to have a non-negligible contribution to their ionization
from AGNs and/or shock heating. Based on the composite spectrum,
we can also probe the weaker He II line, which is another
indicator of AGN activity.

From the He II/H$\beta$ ratio (0.054$\pm$0.006), we can see that
the Offset-AGN sample may have up to $\sim$20\% AGN contribution
to the Balmer emission lines \citep{groves06}. The Offset-SF
sample, on the other hand, shows a lower level of possible AGN
contamination (He II/H$\beta$ of 0.039$\pm$0.002). This
distinction is also suggested by the different host-galaxy
properties of the two samples. The Offset-AGN objects with similar
stellar masses to the Main sample have redder colors, older
stellar populations, lower starburst stellar mass fraction,
smaller H$\alpha$ EWs, and smaller specific star formation rates,
which are similar to host galaxies of AGN \citep{kauffmann03b},
whereas the Offset-SF sample is similar to the Main sample in
terms of these galaxy properties. Furthermore, by construction,
the Offset-SF objects have low [S II]/H$\alpha$ and [O
I]/H$\alpha$ in the range favored for typical star-forming
galaxies.

We note, however, that the Offset-SF sample still has a higher
value of He II/H$\beta$ than the Main sample (0.009$\pm$0.001).
There might be some mixing with the Offset-AGN objects because of
uncertainties in emission-line diagnostic ratios. Indeed, three
objects in the Offset-SF sample have cross-identifications with
$ROSAT$ sources, and two of them have optical spectra indicating
broad Balmer lines. However, $\sim$90\% of the Offset-SF sample
show no evidence of AGNs in terms of $ROSAT$-source
cross-identifications and broad Balmer lines. In addition, we have
examined the He II/H$\beta$ ratios from individual galaxy spectra
in the Offset-SF sample, finding that five objects (including the
three with $ROSAT$ cross-identifications) of the 25 have He
II/H$\beta$ $\sim$ 0.05-0.10, while the remaining 20 objects all
have He II/H$\beta$ $\lesssim$ 0.01, similar to the typical He
II/H$\beta$ of the Main sample. Even the five objects with He
II/H$\beta$ $\sim$ 0.05-0.10 could have other possible ionizing
sources including hot stars and shocks \citep{garnett91}. It is
worth mentioning that nondetection in $ROSAT$ or broad Balmer
lines cannot fully exclude the possibility of AGN contamination,
as the majority of obscured AGNs are completely absorbed in the
soft X-ray range and do not show broad Balmer lines. However, if
the dominant cause of the Offset-SF on the BPT diagram for our
Offset-SF sample was mildly obscured AGN contamination, namely, if
their optical spectra were contaminated by AGN ionization at a
lower level than the Offset-AGN sample, then we would expect them
to exhibit intermediate emission-line diagnostic ratios and
host-galaxy properties, relative to the Main and the Offset-AGN
samples. This is not the case. Instead, the Offset-SF objects have
even higher ionization parameters and electron densities than the
Offset-AGN sample, relative to the Main sample. In addition, as
discussed in \S \ref{sec:sub:comparison}, Offset-SF objects have
host-galaxy properties different from those of typical weak AGNs
in SDSS \citep{kauffmann03b}. Therefore, it appears that AGN
excitation does not provide the dominant cause of the anomalous
emission-line ratios for most of the Offset-SF sample.

\subsubsection{Electron Temperature and Oxygen Abundance}\label{sec:sub:te}

\begin{figure*}
\epsscale{1.} \plotone{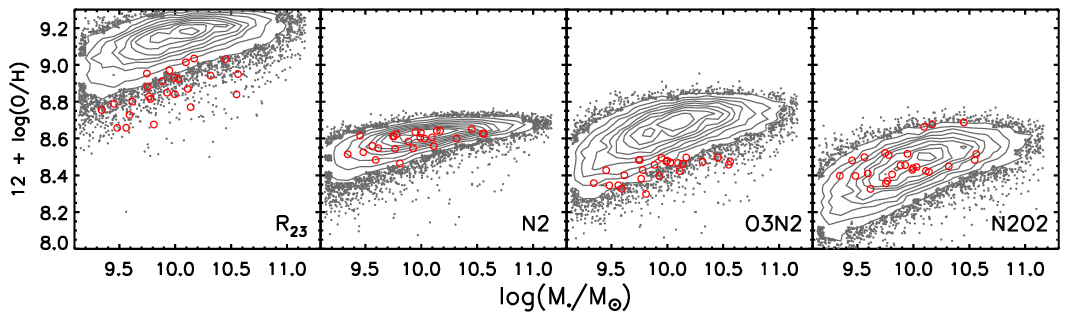} \caption{Mass-metallicity relation
for the SDSS Main ({\it grey contours}) and Offset-SF ({\it red
open circles}) samples, based on various strong-line indicators.
Calibrations are from \citet[][for $R_{23}$]{zaritsky94},
\citet[][for N2 and O3N2]{pp04}, and \citet[][for
N2O2]{bresolin07}, respectively. \label{fig:sdss:massz}}
\end{figure*}

%-------------------------------------------
\begin{deluxetable*}{lcccccc}
\centering
\tabletypesize{\scriptsize}
\tablecaption{Physical Quantities from SDSS composite spectra.\label{tab:sdssabundance}}
\tablewidth{0pt}
\tablehead
{
\colhead{~~~~~~~~~~Sample~~~~~~~~~~} & 
\colhead{~~~~~N$_e$\tablenotemark{a}~~~~~} &
\colhead{~~~~~T$_e$[O III]\tablenotemark{b}~~~~~} &
\colhead{~~~~~T$_e$[O II]\tablenotemark{b}~~~~~} & 
\colhead{~~~~~O$^+$/H$^+$\tablenotemark{c}~~~~~} &
\colhead{~~~~~O$^{++}$/H$^+$\tablenotemark{c}~~~~~} & 
\colhead{~~~~~12 + log(O/H)\tablenotemark{d}~~~~~}
}
\startdata
Main\dotfill      &  28$\pm$ 5 & 0.950$^{+0.124}_{-0.072}$ &  0.965$^{+0.087}_{-0.050}$ & 13.2$^{+3.7}_{-4.1}$ &  3.7$\pm$1.3 & 8.23$^{+0.11}_{-0.17}$ \\
Offset-SF\dotfill & 190$\pm$12 & 1.170$^{+0.039}_{-0.033}$ &  1.119$^{+0.027}_{-0.023}$ &  6.5$\pm$0.6         &  5.2$\pm$0.5 & 8.07$\pm$0.04    \\
\enddata
\tablenotetext{a}{Electron density and 1 $\sigma$ error in units of cm$^{-3}$ derived from $({\rm [S \,\,II]} \lambda 6717/{\rm [S \,\,II]} \lambda 6731$).}
\tablenotetext{b}{Electron temperature and 1 $\sigma$ error in units of $10^4$ K.}
\tablenotetext{c}{Ionic abundance and 1 $\sigma$ error from collisionally excited lines in units of $10^{-5}$.}
\tablenotetext{d}{Total oxygen abundance and 1 $\sigma$ error assuming O/H = (O$^{+}$ + O$^{++}$)/H$^{+}$.}
\end{deluxetable*}
%%%----------------------------------------------------------------------------------

By making composite spectra, and hence enhancing the S/N of weak
auroral lines, we can determine metallicities with a method
independent of strong-line indicators. We have already discussed
the fact that the Offset-SF objects have larger ionization
parameters, as indicated by their higher $O_{32}$ values. However,
it is still unclear whether the Offset-SF sample has comparable
metallicity with the Main one, on average, as suggested by N2 and
N2O2, or if it has $\sim$0.3 dex lower metallicity, as suggested
by O3N2 and $R_{23}$. Settling this question is important for
multiple reasons.

First, gas-phase metallicity and electron temperature, themselves,
serve as key parameters of H~II regions. Second, understanding the
relative metallicities of the Main and the Offset-SF samples on
average can help us determine the difference in average ionization
parameter quantitatively. If the Offset-SF sample has comparable
metallicities with the Main sample, their ionization-parameter
difference would be very large; on the other hand, if the
Offset-SF sample has much lower metallicities than the Main one
does, their ionization-parameter difference would be much smaller,
since both lower metallicity and higher ionization parameter can
cause large $O_{32}$. Finally, if the Offset-SF objects actually
have much lower metallicities than the Main objects, then they
will not follow the local mass-metallicity relation
\citep{tremonti04}, given their comparable stellar masses.

In order to determine oxygen abundances using the direct method,
estimates of both the electron density and temperature are
required. Based on the ratio [S II] $\lambda$6717/[S II]
$\lambda$6731, we establish that on average all of our objects are
in the low-density regime. Then using the five-level atom program
{\tt nebular} implemented in IRAF/STSDAS \citep{shaw95}, we derive
the electron temperature $T$[O III] from the ratio of auroral line
[O III] $\lambda$4363 to nebular lines [O III] $\lambda
\lambda$4959, 5007 \citep{agn2} and correct for dust extinction
using the Balmer decrement method. As for $T$[O II], we adopt a
simple scaling relation between the temperatures in different
ionization zones of an H II region predicted by the
photoionization models of Garnett (1992):
\begin{equation}
T{\rm [O \,\,II]} = 0.70T{\rm [O\,\, III]} + 3000 \,\,{\rm K},
\end{equation}
which is applicable in a wide range of $T$[O III] (2000-18,000 K).
Ionic abundances O$^{++}$/H$^{+}$ and O$^{+}$/H$^{+}$ are then
determined using programs in the {\tt nebular} package, and,
finally, we obtain the total gas phase oxygen abundance assuming
O/H = (O$^{+}$ + O$^{++}$)/H$^{+}$. The results are summarized in
Table \ref{tab:sdssabundance}.

First, the Offset-SF sample has $\sim$2200 K higher electron
temperatures than the Main one does on average. These objects are
unusual in the sense that they have higher electron densities and
electron temperatures, hence ambient interstellar pressures
assuming pressure equilibrium between H II regions and ambient
gas, and larger ionization parameters. It is not clear what is at
the root of the special H II region physical conditions, in terms
of galaxy properties, or larger scale environments. However, in
the simple theoretical estimates for local starburst galaxies, the
ISM pressure scales roughly linearly with the SFR surface density
\citep[e.g.,][]{thompson05}. Indeed, the Offset-SF sample has
several prominently related properties, including higher
concentration ($r$-band Petrosian concentration index
$R_{90}/R_{50}$ of $\sim$2.6 compared to $\sim$2.4), smaller
half-light radii [$R_{50}(z)$ of 1.2 kpc compared to 1.9 kpc], and
most notably, higher SFR surface density
[log($\dot{\Sigma}_{\star}$) of -0.86 M$_{\odot}$ yr$^{-1}$
kpc$^{-2}$ compared to -1.12 M$_{\odot}$ yr$^{-1}$ kpc$^{-2}$)],
as shown in Figure \ref{fig:subfig:dis}, compared to typical
star-forming galaxies. This higher SFR surface density may
therefore account for the higher interstellar pressure seen in the
H II regions of Offset-SF objects.

It is also possible to relate electron temperature, electron
density, and ionization parameter in a single H II region, through
the classical picture of \citet{stromgren39}. With higher ambient
interstellar pressure due to higher electron densities and
temperatures, H II regions are surrounded by denser molecular gas
dust and therefore have smaller radii under the stall condition.
In an idealized case of a fully ionized spherical H II region with
pure hydrogen, the radius $R$ of the ionized region is given by $R
= [3Q/(4\pi N_e^2 \alpha_B)]^{1/3}$, where $Q$ is the rate of
emission of hydrogen ionizing photons, and $\alpha_B$ is the case
B recombination coefficient \citep{agn2}. The ionization parameter
$U$ defined by $[F/4\pi c R^2 N_{{\rm H II}}]$, where $F$ is the
flux of ionizing photons, $c$ is the speed of light, and $N_{{\rm
H II}}$ is the particle density in H II region, is then
proportional to $N_e^{1/3}$ for a given central hot star.
Therefore, in this simple picture, H II regions with higher
electron densities and temperatures will have higher ionization
parameters. Accordingly, the higher electron densities and
temperatures inferred from the integrated spectra of the Offset-SF
objects may result in the observed higher ionization parameters as
well.

Second, the Offset-SF sample has $\sim $0.16 dex lower
metallicities than the Main sample. Note that the absolute
abundance values inferred using the $T_e$ method are significantly
lower than those based on any strong-line indicators. This
difference may stem from temperature fluctuations in H~II regions
and the strong $T_e$ dependence of line emissivities, from which
the net effect is that the electron temperature derived from
auroral lines tends to overestimate the real temperature and hence
systematically underestimate the abundance \citep{garnett92}.
Therefore, we only focus on relative differences based on the same
method between the Offset-SF and Main samples. Given that the
Offset-SF and Main samples are characterized by similar average
stellar masses, this difference in average metallicity suggests
that Offset-SF objects do not follow the local mass-metallicity
relation, in addition to being more compact and sustaining higher
interstellar pressures than the Main sample.

To investigate this question further, we examine several versions
of the mass-metallicity relation for the Offset-SF and the Main
objects, based on different strong-line metallicity indicators.
Figure \ref{fig:sdss:massz} includes these mass-metallicity
correlations for both the Main and the Offset-SF samples,
demonstrating that one would draw different conclusions about the
relationship between the Offset-SF and the Main samples, depending
on which strong-line indicator was used. Between the Main and
Offset-SF samples, the relative average metallicity differences at
fixed mass based on $R_{23}$, N2, O3N2, and N2O2 are $\sim$0.28,
0.00, 0.18, and 0.00 dex, respectively. This ambiguity could be
understood as the result of adopting the same calibration for
samples with different physical conditions, of which the most
important are ionization parameter and electron density.

The discrepancy between conclusions drawn from the strong-line
indicators and the $T_e$ method can be understood using the
photoionization models of \citet{kewley02}. For example, according
to $R_{23}$, the average metallicity difference between the
Offset-SF and Main samples is $\sim$0.28 dex, whereas this
metallicity difference based on the $T_e$ method is $\sim$0.16
dex. This $\sim$0.12 dex discrepancy can be explained by examining
the theoretical grids in the $R_{23}$ versus 12 + log(O/H) space,
given by photoionization models with different values of electron
density and ionization parameter. If the incorrect assumption is
adopted that both the Main and Offset-SF samples have electron
densities $N_e \sim 10$ cm$^{-3}$ and ionization parameters
$q=3\times10^{7}\mbox{cm s}^{-1}$ -- values appropriate for the
Main but not Offset-SF sample --  we obtain that the Offset-SF
sample is $\sim 0.25$~dex lower in oxygen abundance than the Main
sample. On the other hand, if an electron density $N_e \sim 350$
cm$^{-3}$ and ionization parameter $q=5\times10^{7}\mbox{cm
s}^{-1}$ are assumed for the Offset-SF sample --  more appropriate
based on their inferred H~II region physical conditions -- the
difference in estimated Main and Offset-SF oxygen abundance is
only $\sim$0.10 dex. This difference is $\sim$0.15 dex smaller
than the one inferred with the incorrect assumption that the
Offset-SF and Main samples have the same electron densities and
ionization parameters. The actual difference in electron density
between the two samples is less extreme ($N_e \sim 190$ cm$^{-3}$
of the Offset-SF compared to $N_e \sim 28$ cm$^{-3}$ of the Main),
so this discrepancy may be smaller, as observed ($\sim$0.12 dex).
Analogous effects occur for the N2 and O3N2 indicators.

In summary, the metallicities of the Offset-SF objects based on
strong-line indicators can be either overestimated or
underestimated, depending on the relative shifts of the
theoretical grids caused by varying physical conditions, and an
incorrect assumption of the values of ionization parameter and
electron density. For the SDSS Offset-SF sample, the metallicities
based on strong-line indicators can either be systematically too
large by $\sim$0.16 dex (N2, N2O2), or too low by $\sim$0.12 dex
($R_{23}$). The bias of O3N2 is little (too large by $\sim$0.02
dex) for our SDSS Offset-SF objects, although it may not always be
negligible given other ranges of physical conditions in terms of
metallicity, ionization parameter, and electron density. The
reason for the Offset-SF objects' anomalous positions on the BPT
diagram is the same: H~II regions with distinct ionization
parameters and electron densities fall on different surfaces in
the diagnostic-line parameter space.

One caveat that should be mentioned here is the fact that we are
inferring average biases over the stellar mass and metallicity
ranges spanned by the Offset-SF sample. Another caveat is that
results of the SDSS data are all based on integrated fiber spectra
and hence limited in terms of spatial information. We plan to
further study the spatial dependence of H~II region emission lines
in these offset objects using long-slit spectrographs in the
future.

\subsection{Implications for DEEP2 Objects}

We have analyzed emission-line diagnostic ratios, physical
conditions and galaxy properties of SDSS objects with similar [N
II]/H$\alpha$ and [O III]/H$\beta$ values to those of DEEP2
galaxies with the most extreme offset in our sample, and $z \sim
2$ star-forming galaxies \citep{erb06a}. There are two major
causes for their offset: one is different H II region physical
conditions characterized by higher electron density and
temperature, and hence larger ionization parameter, compared to
typical SDSS star-forming galaxies. These physical conditions of
the SDSS Offset-SF objects are also connected to their host galaxy
properties, particularly the higher SFR surface density. The other
possible cause for the offset is contribution from AGNs and/or
shock excitation. We cannot rule out either of these two
possibilities, or a combination of both, to explain the offset of
DEEP2 objects on the diagnostic diagram.

As for the question of H II region physical conditions, it is
possible to determine electron densities for our DEEP2 objects
using the [O II] doublet contained in the DEIMOS spectra. Note
that we will not directly compare electron densities of the $z
\sim 1.0-1.5$ objects to those of SDSS local galaxies, because
they were estimated using different density-sensitive doublets,
and therefore the comparison might suffer from the associated
systematic uncertainties. However, when we divide all of the
objects in our DEEP2 sample that have reliable [O II]-doublet
measurements into two groups, according to the empirical Ka03
curve in the [N II]/H$\alpha$ versus [O III]/H$\beta$ diagram, we
find that for the group below the Kauffmann curve, the median
value of electron density inferred from [O II] $\lambda$3726/[O
II] $\lambda$3729 is $\sim$23 cm$^{-3}$, whereas for the group
above the Kauffmann curve, the median value is $\sim$159
cm$^{-3}$. This difference in electron density as a function of
position on the BPT diagram is qualitatively consistent with what
we have found for SDSS objects (Table (\ref{tab:sdssabundance});
electron density N$_e$ [cm$^{-3}$] of 190$\pm$12 for the Offset-SF
compared to 28$\pm$5 for the Main), lending independent support
for our method of using Offset-SF objects as local analogs for the
H~II regions in our high-redshift sample. In principle, we could
examine the ionization parameter indicator, $O_{32}$, for our
DEEP2 objects, using [O II] from DEIMOS and [O III] from NIRSPEC
spectra. However, at this point, the comparison does not seem
useful, as it would suffer from large systematic uncertainties,
due to the manner in which the optical and near-IR spectra were
collected and flux-calibrated.

A related question is whether objects that are more offset in the
[N II]/H$\alpha$ versus [O III]/H$\beta$ diagram have higher
average SFR surface density than objects with less offset within
our DEEP2 $z \sim 1-1.5$ sample. The spatial extent of H$\alpha$
emission contains information about galaxy sizes. However, we can
only measure the extent along the slit, which does not necessarily
reflect the actual galaxy size. In addition, the size estimated
from the spatial extent of H$\alpha$ emission is subject to
uncertainties due to the variation in the seeing FWHM, which is
comparable to the H$\alpha$ size itself. As we do not have
reliable estimates for the sizes of our DEEP2 objects, the above
question can be finally answered only when the galaxy
morphological information is robustly gathered. Furthermore, the
SFRs of our DEEP2 objects have not been corrected for dust
extinction or aperture effects, which may amount to factor of 2
differences \citep{erb06c}. Limited by the associated systematic
uncertainties, we are not able to compare the SFR surface
densities of our DEEP2 objects directly with those of the SDSS
samples. However, after correcting the SFR-associated systematic
uncertainties for their $z \sim 2$ sample, \citet{erb06b} found a
mean log($\dot{\Sigma}_{\star}$) $\approx$ 0.46, significantly
higher than that of SDSS local star-forming galaxies. In addition,
observational evidence exists that the galaxy size at fixed
mass/luminosity decreases with increasing redshift out to z $\sim$
3 \citep{trujillo06,dahlen07}. At the same time, galaxies with
star formation rates significantly higher than those found among
typical SDSS emission-line galaxies are more commonly found at
higher redshifts \citep[e.g.,][]{dahlen07,reddy07}. Therefore,
galaxies with high SFR surface densities are more prevalent in the
high-redshift universe. For such objects, a noticeable difference
in H~II region physical conditions is expected.

As for the question of AGN contribution, the DEEP2 object 42010637
clearly falls on the part where many objects have contribution
from AGNs in their ionizing spectra. However, it does not have
multi-wavelength information that can either confirm or rule out
the AGN excitation. In addition, all of the $z\sim 2$ objects with
[O III]/H$\beta$ and [N II]/H$\alpha$ measurements are offset by
an amount similar to our most extremely offset $z\sim 1.0-1.5$
objects but show no evidence of AGN contamination in at least
their rest-frame UV spectra \citep{erb06a}. The lack of such
evidence in the rest-frame UV rules out the presence of at least a
fairly unobscured AGN. Note that \citet{daddi07} find that roughly
20-30\% of star-forming galaxies with $M\sim 10^{10}$-$10^{11}
M_{\odot}$ at $z\sim 2$ display a mid-IR excess, as evidence for
hosting an obscured AGN. These authors show that this fraction
increases with stellar mass, reaching $\sim$50-60\% for an extreme
mass range of $M > 4\times10^{10} M_{\odot}$. Our DEEP2 $z\sim
1.0-1.5$ sample covers a range of $M \sim 5\times10^9$- $10^{11}
M_{\odot}$, not all of which are in the ``massive'' range of
\citet{daddi07}, and only two of 20 are in the extreme range. If
the mid-IR excess, hence AGN contamination, found by
\citet{daddi07} is also applicable to our DEEP2 $z\sim 1.0-1.5$
objects, the fraction of objects containing potential AGN
contamination would be at most $\sim 20$\%-30\%. However, nine (or
10, one with a lower limit) of the 13 objects in our sample with
all four line measurements show evidence for being offset in the
[O III]/H$\beta$ versus [N II]/H$\alpha$ BPT diagram (i.e., more
than 20-30\%).  More generally, the associated space densities of
both the blue DEEP2 galaxies \citep{coil07} and UV-selected $z
\sim 2$ objects for which we have NIRSPEC spectra
\citep{adelberger05,reddy07} are significantly higher than that of
the obscured AGN population featured in \citet{daddi07}.
Therefore, as one of the possible causes for the offset on the BPT
diagram, AGN contamination cannot account for nor is consistent
with all the rest-frame optical emission-line measurements
presented here.

We will address the AGN-contamination issue in the future by
looking at more DEEP2 objects for which multi-wavelength
information is available and obtaining spatially resolved spectra
with an integral field unit assisted by adaptive optics to isolate
the contribution from the nucleus. Also, the measurements of
flux-calibrated emission lines including [O II] $\lambda$3727, [O
III] $\lambda$4363, H$\beta$, [O III] $\lambda$5007, [O I]
$\lambda$6300, [N II] $\lambda$6584, H$\alpha$ and [S II]
$\lambda\lambda$6717, 6731, as well as host-galaxy morphological
information are required, in order to finally settle the causes of
the offset in the [N II]/H$\alpha$ versus [O III]/H$\beta$ diagram
for high-redshift star-forming galaxies. [O III] $\lambda$4363 may
be difficult to measure for all but the most metal-poor objects or
in deep $z \sim 1$ composite spectra, and intensity limits on [O
I] $\lambda$6300 may also be difficult to obtain. Yet measurements
of the stronger emission lines for a statistical sample of objects
at $z \geq 1$ will be feasible with the next generation of
ground-based multi object near-IR spectrographs.

In an independent study, \citet{brinchmann07} have also analyzed
the possible causes for the high-redshift galaxies' offset in the
BPT diagram, using theoretical models of nebular emission from
star-forming galaxies \citep{charlot01} and the SDSS DR4 data.
They have found a relationship in SDSS galaxies between their
location in the BPT diagram and their excess specific SFRs and
larger H$\alpha$ EWs relative to galaxies of similar mass. We note
that they have only examined SDSS star-forming galaxies below the
Ka03 curve, whereas our SDSS Offset samples have been selected to
be above this curve, in order to probe exactly the regime where
the most offset objects in our DEEP2 $z \sim 1-1.5$ sample and the
\citet{erb06a} $z - 2$ star-forming galaxies reside.
\citet{brinchmann07} have inferred that an elevated ionization
parameter $U$ is at the root of the excess specific SFRs of the
more offset objects within their pure star-forming galaxy sample
and further speculated that higher electron densities and escape
fractions of hydrogen ionizing photons might be the factors
responsible for the systematically higher values of $U$ in the H
II regions of high-redshift galaxies. Using a different technique
and sample of galaxies, we have reached a similar conclusion about
the higher ionization parameter and larger H$\alpha$ EWs in our
SDSS Offset-SF samples. In addition, we have also uncovered that
the higher electron density and temperature, hence higher
interstellar ambient pressure, is at the root of the higher
ionization parameter. We have further shown that these unusual H
II region physical conditions are well connected to the higher SFR
surface density of host galaxies. The trend of higher electron
density with increasing BPT diagram offset found within our DEEP2
$z \sim 1-1.5$ sample, and the observational evidence that
galaxies with high SFR surface densities are more prevalent at
high redshifts, lend further support to our conclusions drawn
based on the SDSS local emission-line galaxies.

These differences discovered in H II region physical conditions,
which may commonly apply to $z \sim 1-1.5$ star-forming galaxies,
must be taken into account when strong-line abundance indicators
are used to study the evolution of galaxy metallicity with
redshift. The resulting systematic bias in inferred oxygen
abundance can be estimated quantitatively either via detailed
photoionization models, given the difference in H II region
physical conditions inferred from the relevant density- and
ionization parameter-sensitive line ratios, or through empirical
comparisons, as we have illustrated for the SDSS Main and
Offset-SF samples.

We have shown in \S \ref{sec:sub:te} that, for the SDSS Offset-SF
objects, the metallicities based on various strong-line indicators
can either be systematically too large by $\sim$0.16 dex (N2,
N2O2) or too low by $\sim$0.12 dex ($R_{23}$). For N2 in
particular, the inferred metallicities of the SDSS Offset-SF
objects can be systematically too large by $\sim$0.16 dex, which
is already comparable to the inherent scatter in the N2
calibration \citep[1 $\sigma$ dispersion of 0.18 dex;][]{pp04}.
This systematic bias may be even larger if the actual difference
in H II region physical conditions is more extreme. In addition,
we note that this systematic uncertainty stemming from the
difference in H II region physical conditions has different
effects from that of the inherent scatter in the calibration, when
the strong-line relation calibrated with local H II regions is
applied to high-redshift star-forming galaxies. As there is
evidence that the average ionization parameter and electron
density in high-redshift star-forming galaxies are systematically
higher than the local typical values, there will be a systematic
``bias,'' instead of a scatter as a result. For example, the
N2-based metallicities for the DEEP2 $z\sim 1.4$ sample may be
systematically overestimated by as much as $\sim$ 0.16 dex.
Furthermore, as discussed in \S 4, the fact, that our DEEP2 $z\sim
1.4$ sample is more offset from the local excitation sequence than
the $z\sim 1.0$ sample, may result in N2-based metallicities that
are more significantly overestimated as well. This effect may
cause the apparent reverse trend of average O/H with redshift
within the DEEP2 sample, such that the $z\sim 1.4$ sample is
described by apparently higher metallicities at fixed stellar mass
than the one at $z\sim 1.0$, despite the general trend of
increasing O/H towards lower redshift. The reverse trend within
the DEEP2 sample based on O3N2 is less significant than that based
on N2, which is consistent with the fact that the O3N2-based
metallicity bias due to the offset in the BPT diagram is much less
than that of N2.

In addition, we can quantify the potential bias for the sample of
star-forming galaxies at $z\sim2$ presented by \citet{erb06a}. If
the four $z\sim2$ objects with the full set of H$\beta$, [O III]
$\lambda 5007$, H$\alpha$, and [N II] $\lambda 6584$ measurements
are representative of the larger population of star-forming
galaxies in \citet{erb06a} in terms of diagnostic line ratios,
such that the $z\sim 2$ star-forming galaxies are even more
significantly offset than the DEEP2 $z\sim 1.4$ sample on average,
then their metallicities based on N2 would also be overestimated
by $\sim$0.16 dex. Therefore, the true metallicity offset between
the  \citet{erb06a} $z\sim2$ sample and local SDSS objects may be
as much as $\sim$0.16 dex larger than it appears now in Figure
(\ref{fig:massz}a). Accounting for the systematic differences in
converting strong emission-line ratios to oxygen abundances is
therefore a crucial component of comparing galaxy metallicities at
different redshifts.

\section{Summary}

We have compiled a sample of 20 star-forming galaxies at $1.0 < z
< 1.5$ drawn from the blue cloud of the color bimodality observed
in the DEEP2 survey, to study the correlation between stellar mass
and metallicity, across a dynamical range of 2 orders of magnitude
in stellar mass, as well as H~II region physical conditions at
this redshift range. In order to gain some insights on the causes
of the offset in the BPT diagram observed in high-redshift
star-forming galaxies, we have examined the H II region diagnostic
line ratios and host galaxy properties of the small fraction of
SDSS galaxies that have similar diagnostic ratios to those of the
DEEP2 sample. Our main results are summarized as follows:
\begin{enumerate}

\item[1.] There is a correlation between stellar mass and
gas-phase oxygen abundance in DEEP2 star-forming galaxies at $z
\sim 1.0$ and at $z \sim 1.4$. We have found that the zero point
of the $M-Z$ relationship evolves with redshift, in the sense that
galaxies at fixed stellar mass become more metal-rich at lower
redshift, by comparing the $1.0 < z < 1.5$ sample with UV-selected
$z \sim 2$ and SDSS local star-forming galaxies. At the low-mass
end ($M_{\star} \sim 8\times10^9 M_{\odot}$), the relation at $1.0
< z < 1.5$ is offset by $\sim$0.2 (0.35) dex from the local
mass-metallicity relation according to the N2 (O3N2) indicator.
The N2-based offset could be larger by as much as $\sim$0.16 dex,
when the systematic bias due to difference in H II region physical
conditions between $1.0 < z < 1.5$ and the local universe is taken
into account. At the high-mass end ($M_{\star} \sim 5\times10^{10}
M_{\odot}$), the metallicity offset between the DEEP2 $1.0 < z <
1.5$ sample and the local SDSS sample is at least $\sim 0.2$ dex,
according to the O3N2 indicator.

\item[2.] As observed previously for a very small sample of
high-redshift galaxies, on average our new DEEP2 sample at $1.0 <
z < 1.5$ is offset from the excitation sequence formed by nearby
H~II regions and SDSS emission-line galaxies. By examining the
small fraction of SDSS galaxies that have similar diagnostic
ratios to those of the DEEP2 sample, we have found two likely
causes for the anomalous emission-line ratios. One is the
contribution from AGN and/or shock excitation at the level of
$\sim 20$\%. The other is the difference in H~II region physical
conditions, characterized by significantly larger ionization
parameters, as a result of higher electron densities and
temperatures, and hence higher interstellar ambient pressure, than
the typical values of local star-forming galaxies with similar
stellar mass. These unusual physical conditions are possibly
connected to the host-galaxy properties, most importantly smaller
sizes and higher star-formation rate surface densities. Our
conclusion drawn from analyzing the SDSS data has been further
verified by the fact that the DEEP2 objects more offset from the
local excitation sequence in the BPT diagram also have higher
electron densities than those closer to the local sequence. We
cannot rule out either the contribution from AGN and/or shock
excitation, or the difference in H~II region physical conditions,
for the unusual emission-line diagnostic ratios of high-redshift
star-forming galaxies.

\item[3.] We have quantified the effects of different H II region
physical conditions on the strong-line metallicity calibrations.
The direct electron temperature method was used to estimate the
``true'' metallicity difference between offset SDSS objects with
anomalous line ratios and more typical objects of similar stellar
mass. Strong-line indicators were also used to estimate this
difference. A comparison of these results reveals potential biases
in the strong-line indicators. According to our test, the
metallicities based on strong-line indicators can either be
systematically too large by $\sim$0.16 dex (N2, N2O2), or too low
by $\sim$0.12 dex ($R_{23}$), for objects with similar H~II region
physical conditions to those observed in high-redshift galaxies.
The bias of O3N2 is much less significant (too large by $\sim$0.02
dex) for offset SDSS objects with anomalous line ratios, although
it may not always be negligible given other ranges of physical
conditions in terms of metallicity, ionization parameter, and
electron density.
\end{enumerate}

The difference in H~II region physical conditions, which may
commonly apply to $z\sim 1.0-1.5$ star-forming galaxies, must be
taken into account when strong-line abundance indicators are used
to study the evolution of galaxy metallicity with redshift. There
are at least two methods to remove the systematic bias from the
effect of significantly different H~II region physical conditions
on the strong-line abundance calibrations. One is to gather the
abundance information with direct $T_e$ method for a sample of
high-redshift H~II regions as the calibration sample, which may be
hard to achieve, as auroral lines are difficult to measure except
for very metal-poor objects or in deep, composite spectra. The
other, which is currently feasible yet relies on photoionization
models, is to quantify the biases in strong-line indicators when
certain physical conditions are present and then compensate the
biases when inferring abundances from strong-line indicators for
high-redshift galaxies.

In this study we have presented evidence that high-redshift
star-forming galaxies possess distinct H~II region physical
properties, as characterized by on average larger ionization
parameters, higher electron densities, and temperatures, which are
possibly connected to their relatively smaller sizes and higher
SFR surface densities. These conditions may be quite common during
the epoch at $z \geq 1$ when at least 50\% of the local stellar
mass density was formed \citep{bundy06,drory05}. Therefore, they
should be characterized in more detail for a full understanding of
the star formation history of the universe as well as the buildup
of heavy elements in galaxies. The next generation of ground-based
near-IR multi-object spectrographs will play a key role in
assembling rest-frame optical emission-line measurements for large
samples of high-redshift galaxies, enabling the detailed study of
star-forming galaxies in the early universe.

\acknowledgments We are indebted to the DEEP2 team, whose
significant efforts in establishing such a tremendous
spectroscopic sample at $z\sim1$ made this project possible. We
also thank Kevin Bundy for his assistance with estimating stellar
masses, and Bruce Draine and Jenny Greene for helpful discussions.
A. E. S. acknowledges support from the David and Lucile Packard
Foundation and the Alfred P. Sloan Foundation. A. L. C. is
supported by NASA through Hubble Fellowship grant HF-01182.01-A
awarded by the Space Telescope Science Institute, which is
operated by the Association of Universities for Research in
Astronomy, Inc., for NASA, under contract NAS 5-26555. C. P. M. is
supported in part by NSF grant AST 04-07351. Funding for the DEEP2
survey has been provided by NSF grants AST95-09298, AST-0071048,
AST-0071198, AST-0507428, and AST-0507483 as well as NASA LTSA
grant NNG04GC89G. Funding for the Sloan Digital Sky Survey (SDSS)
has been provided by the Alfred P. Sloan Foundation, the
Participating Institutions, the National Aeronautics and Space
Administration, the National Science Foundation, the U.S.
Department of Energy, the Japanese Monbukagakusho, and the Max
Planck Society. The SDSS Web site is http://www.sdss.org/. We wish
to extend special thanks to those of Hawaiian ancestry on whose
sacred mountain we are privileged to be guests.  Without their
generous hospitality, most of the observations presented herein
would not have been possible.

\bibliographystyle{apj}
\bibliography{apj-jour,z1sfgrefs}

\end{document}